\documentclass[12pt]{article}
\usepackage{amsmath}
\usepackage{amssymb}
\begin{document}
\begin{center}
{\bf {\large{Supervariable and BRST Approaches to a \\ Reparameterization Invariant Non-Relativistic System}}}

\vskip 2.5cm

{\sf  A. K. Rao$^{(a)}$, A. Tripathi$^{(a)}$, R. P. Malik$^{(a,b)}$}\\
$^{(a)}$ {\it Physics Department, Institute of Science,}\\
{\it Banaras Hindu University, Varanasi - 221 005, (U.P.), India}\\

\vskip 0.1cm

$^{(b)}$ {\it DST Centre for Interdisciplinary Mathematical Sciences,}\\
{\it Institute of Science, Banaras Hindu University, Varanasi - 221 005, India}\\
{\small {\sf {e-mails: amit.akrao@gmail.com; ankur1793@gmail.com; 
  rpmalik1995@gmail.com}}}
\end{center}

\vskip 1.5 cm

\noindent
{\bf Abstract:}
We exploit the theoretical strength of the supervariable and Becchi-Rouet-Stora-Tyutin (BRST) formalisms to derive the proper (i.e. off-shell
nilpotent and absolutely anticommuting) (anti-)BRST symmetry transformations for the reparameterization invariant model of a non-relativistic (NR)
free particle whose space $(x)$ and time $(t)$ variables are function of an evolution parameter $(\tau)$. The infinitesimal 
reparameterization (i.e. 1D diffeomorphism) symmetry transformation of our theory is defined w.r.t. {\it this} evolution parameter $(\tau)$. 
We apply the {\it modified} Bonora-Tonin (BT) supervariable approach (MBTSA) as well as the (anti-)chiral supervariable approach 
(ACSA) to BRST formalism to discuss various aspects of our present system. For this purpose, our 1D {\it ordinary} theory (parameterized by $\tau$) is generalized 
onto a $(1, 2)$-dimensional supermanifold which is characterized by the superspace coordinates $Z^M = (\tau, \theta, \bar\theta)$ where
a pair of Grassmannian variables satisfy the {\it fermionic} relationships: $\theta^2 = {\bar\theta}^2 = 0, \, \theta\,\bar\theta + \bar\theta\,\theta = 0$
and $\tau$ is the {\it bosonic} evolution parameter. In the context of ACSA, we take into account {\it only} the (1, 1)-dimensional (anti-)chiral super 
sub-manifolds of the {\it general} (1, 2)-dimensional supermanifold. The derivation of the {\it universal} Curci-Ferrari (CF)-type restriction, from various 
underlying theoretical methods, is a novel observation in our present endeavor.
Furthermore, we note that the {\it form} of the gauge-fixing and Faddeev-Popov ghost terms for our NR and non-SUSY system is exactly {\it same} as {\it that} 
of the reparameterization invariant SUSY (i.e. spinning) and non-SUSY (i.e. scalar) {\it relativistic} particles. This is a novel
observation, too.



\newpage

\section {Introduction}

\noindent
During the last few years, there has been an upsurge of interest in the study of diffeomorphism invariant theories because 
one of the key and decisive features of the gravitational and (super)string theories is the observation that they respect the 
{\it classical} diffeomorphism symmetry transformations. The {\it latter} symmetry transformations can be exploited within the 
framework of Becchi-Rouet-Stora-Tyutin (BRST) formalism [1-4] where the {\it classical} diffeomorphism symmetry transformation 
is elevated to the {\it quantum} (anti-)BRST symmetry transformations. 
In fact, it is the key feature of the BRST formalism that the {\it classical} diffeomorphism transformation {\it parameter} is traded 
with the fermionic (anti-)ghost fields/variables at the {\it quantum} level. In other words, the (anti-)BRST 
transformations are of the supersymmetric (SUSY) kind under which the bosonic type of fields/variables transform to the fermionic type fields/variables and 
vice-versa. Two of the key properties of the (anti-)BRST transformations are the on-shell/off-shell nilpotency and absolute 
anticommutativity. These key properties encompass in their folds the {\it fermionic} as well as {\it independent} natures of the {\it quantum} 
BRST and anti-BRST symmetries at the level of {\it physical} interpretation. The nilpotency property (i.e. {\it fermionic} nature) of 
the (anti-)BRST symmetries (and their corresponding charges) is {\it also} connected with some aspects of the cohomological properties of 
differential geometry and a few decisive features of supersymmetry.

The BRST formalism has been exploited in the covariant canonical {\it quantization} of the {\it gauge} and {\it diffeomorphism} 
invariant theories in the past. At the {\it classical} level, the {\it gauge theories} are characterized by the existence of the first-class constraints [5, 6]
{\it on} them. This fundamental {\it feature} is translated, at the {\it quantum} level, into the language of the existence of the Curci-Ferrari (CF)-type 
restriction(s) when the {\it classical} theory is quantized by exploiting the theoretical richness of BRST formalism. Hence, the existence of the CF-type 
restriction(s) is the key signature of a BRST-{\it quantized} version of the gauge and/or diffeomorphism invariant theory. The CF-type restrictions
are (i) deeply connected with the geometrical objects called gerbes [7, 8], (ii) responsible for the absolute anticommutativity of the {\it quantum} (anti-)BRST 
transformations, and (iii) the root-cause behind the existence of the coupled (but equivalent) Lagrangians/Lagrangian densities for the
(anti-)BRST invariant {\it quantum} theories (corresponding to the {\it classically} gauge/diffeomorphism invariant theories).
The Abelian 1-form gauge theory is an {\it exception} where we obtain a {\it unique} (anti-) BRST invariant Lagrangian density because the 
CF-type restriction is {\it trivial} in this case. However, {\it this} restriction turns out to be the 
limiting case of the non-Abelian 1-form gauge theory where the {\it non-trivial} CF-condition exists [9].

It is the supervariable/superfield approaches [10-21] to  BRST formalism that provide the geometrical basis for the off-shell 
nilpotency and absolute anticommutativity of the (anti-) BRST symmetries as well as the existence of the CF-type restrictions for 
a BRST-{\it quantized} gauge/diffeomorphism invariant theory.  In the usual superfield approaches (USFA), it is the horizontality condition (HC)
that plays a decisive role as it leads to (i) the derivation of the (anti-)BRST symmetry transformations for {\it only} the gauge and anti-)ghost fields,
as well as (ii) the derivation of the CF-type restriction(s). The augmented version of the superfield approach (AVSA) is an extension of USFA 
where, in addition to the HC, the gauge  [i.e. (anti-)BRST] invariant restrictions are exploited {\it together} which lead to the derivation
of the (anti-)BRST symmetry transformations for the gauge, (anti-)ghost and {\it matter} fields {\it together} in an {\it interacting} gauge theory. 
It has been a challenging problem to incorporate 
the diffeomorphism transformation within the ambit of superfield approach to gauge theories (see, e.g. [14-16]) so that one can discuss the gravitational 
and (super)string theories within the framework of USFA/AVSA. In this direction, a breakthrough has recently been made by Bonora [22] where the superfield 
approach has been applied to derive the proper (anti-)BRST transformations as well as the CF-type restriction for the D-dimensional 
diffeomorphism invariant theory. This approach has been christened by us as the {\it modified} Bonora-Tonin (BT) superfield approach (MBTSA) 
to BRST formalism. In a recent couple of papers [23, 24], we have applied the theoretical beauty of the MBTSA as well as ACSA [i.e. 
(anti-)chiral superfield/supervariable approach] to BRST formalism [25-29] in the context of the 1D diffeomorphism (i.e. reparameterization) invariant theories of
the non-SUSY (i.e. scalar) as well as SUSY (i.e. spinning) {\it relativistic} free particles.

The central theme of our present investigation is to concentrate on the reparameterization (i.e. 1D diffeomorphism) invariant theory of a massive
non-supersymmetric (NSUSY) and non-relativistic (NR) free particle where the {\it standard} NR Lagrangian $L_{0}^{(t)} (x, \dot x) 
= \frac{1}{2} m \, {\dot x}^2$ 
(with $ \dot x = d x / d t)$ is rendered reparameterization invariant by treating the ``time" variable on a par with $x$ variable [30] 
that are parameterized by an evolution parameter $\tau$
such that the {\it new} Lagrangian $L_{0}^{(\tau)} (x,\, \dot x, \, t, \, \dot t) = \frac{m\,{\dot x}^2}{2\,\dot t} $ where $\dot x = d x / d \tau$
and $\dot t = d t/ d \tau$. The {\it latter} Lagrangian respects the reparameterization symmetry [31, 32] and it has been discussed in different 
theoretical settings
where the noncommutativity of the spacetime appears by the symmetry considerations, constraints analysis, redefinitions of variables, etc. 
This reparameterization invariant model of the free particle ($\dot p_{x} = \dot p_{t} = 0$) has been discussed by us within the 
frameworks of BRST formalism as well as quantum groups [32]. However, in the BRST analysis, we have exploited the gauge symmetry of 
this NSUSY and NR system [32] {\it without} discussing anything about the reparameterization transformations. In our present 
investigation, we have {\it applied} the beautiful blend of theoretical ideas from MBTSA and ACSA to derive the proper (anti-)BRST 
symmetries and CF-type restriction for {\it this} NR system. This model is interesting in its own right as it is a 
NR system (unlike our earlier discussions [23, 24] on the {\it relativistic} systems) and ``time" itself has been treated as a physical 
observable that depends on the evolution parameter $\tau$. The {\it latter} property of our present NR system is important as ``time" 
has {\it also} been treated as an {\it observable} in quantum mechanics instead of an evolution parameter (see, e.g. [30]).

The following motivating factors have been at the heart of our curiosity in pursuing our present endeavor. First, so far, we 
have been able to apply the beautiful blend of theoretical ideas behind MBTSA and ACSA to BRST formalism in the cases of
reparameterization invariant systems of (i) relativistic non-supersymmetric (NSUSY) scalar free particle, and (ii) supersymmetric
(SUSY) (i.e. spinning) relativistic free particle. Thus, it has been a challenging problem for us to apply the 
{\it same} theoretical ideas to discuss the NSUSY and non-relativistic (NR) system of a reparameterization invariant free particle. 
We have accomplished this goal in our present investigation.
Second, we have shown the {\it universality} of the CF-type restriction in the cases of reparameterization 
(i.e. 1D diffeomorphism) invariant NSUSY as well as SUSY systems of the free relativistic particles. Thus, we have been motivated to 
see the existence of the {\it same} CF-type of restriction in our present case of reparameterization invariant system of NSUSY 
and NR free particle. We have been able to demonstrate
that it is the {\it same} CF-type of restriction that exists in the BRST approach to our present NSUSY and NR system. Third, 
we have found out that the gauge-fixing and Faddeev-Popov (FP) ghost terms for the systems of non-SUSY (i.e. scalar) and SUSY (i.e. spinning)
relativistic particles are {\it same} within the ambit of BRST formalism. Thus, we have been curious to find out the gauge-fixing and FP-ghost terms for our 
present non-SUSY and NR system. It is surprising that the above {\it terms} are {\it same} for our present system, too. Finally, our present 
work is our {\it another} modest initial step towards our main goal of applying the theoretical potential of MBTSA and ACSA to the physical four 
(3+1)-dimensional (4D) diffeomorphism invariant gravitational and (super)sting theories in the higher dimensions (i.e. $D > 4$) of spacetime.

The contents of our present endeavor are organized as follows. In Sec. 2, we recapitulate the bare essentials of the Lagrangian formulation
of our reparameterization invariant non-relativistic system and discuss the BRST quantization of this model by exploiting {\it its} infinitesimal
{\it gauge} {\it symmetry} transformations. Our Sec. 3 is devoted to the application of MBTSA for the derivation of (i) the {\it quantum} (anti-)BRST symmetry
transformations for the phase space variables, and (ii) the underlying (anti-)BRST invariant Curci-Ferrari (CF)-type restriction (corresponding to the 
{\it classical} infinitesimal reparameterization symmetry transformations). The theoretical content of Sec. 4  is related to the derivation of the {\it full}
set of (anti-)BRST symmetry transformations by requiring the off-shell nilpotency and absolute anticommutativity properties. We {\it also} show the existence
of the CF-type restriction and deduce the coupled (anti-)BRST invariant Lagrangians for our theory. In Sec. 5, we derive the (anti-)BRST symmetry
transformations for the {\it other} variables (i.e. different from the phase variables) within the purview of ACSA. Our Sec. 6 deals with the proof of the 
{\it equivalence} of  the coupled Lagrangians within the framework of ACSA to BRST formalism. In Sec. 7, we prove the off-shell nilpotency and 
absolute anticommutativity of the conserved (anti-)BRST charges in the {\it ordinary} space and superspace (by exploiting the theoretical richness of ACSA to 
BRST formalism). Finally, in Sec. 8, we discuss our key results and point out a few future directions for further investigation(s).

In our Appendices A, B and C, we perform some explicit computations that supplement as well as corroborate the key claims that have been made in the main
 body of our text.\\

\noindent
{\it Convention and Notations}: We adopt the convention of the left-derivative w.r.t. {\it all} the fermionic variables of our theory. In the whole body of
the text, we denote the fermionic (anti-)BRST transformations by the symbols $s_{(a)b}$ and corresponding conserved and nilpotent charges carry the
notations $Q_{(a)b}$ and $Q_{(\bar B)B}$ in different contexts. The general $(1, 2)$-dimensional supermanifold is parameterized by $Z^M = (\tau, 
\theta, \bar\theta)$ and its {\it chiral} and {\it anti-chiral} super sub-manifolds are characterized by $(\tau, \theta)$ and $(\tau, \bar\theta)$,
respectively, where the {\it bosonic} coordinate is represented by the evolution parameter $(\tau)$ and the Grassmannian variables $(\theta, \bar\theta)$
obey the {\it fermionic} relationships: $\theta^2 = 0,\,{\bar\theta}^2 = 0,\,\theta\,\bar\theta + \bar\theta\,\theta = 0$.

\section {Preliminaries: Lagrangian Formulation}

Our present section is divided into {\it two} parts. In subsection 2.1, we discuss the {\it classical} infinitesimal reparameterization and gauge symmetry
transformations. Our subsection 2.2 deals with the BRST quantization of  our system by exploiting the {\it classical} gauge symmetry transformations 
(which are {\it also} infinitesimal and continuous).

\subsection {Classical Infinitesimal Symmetries}

We begin with the {\it three} equivalent reparameterization invariant Lagrangians for the free non-relativistic and non-SUSY 
particle as (see e.g. [32] for details) 
\begin{eqnarray}
&&L_0(x,\,\dot x, \, t, \dot t) = \frac{m\,{\dot x}^2}{2\,\dot t},  \nonumber \\
&&L_f(x,\,\dot x, \, t, \dot t,\,p_x,\,p_t) = p_x\,\dot x + p_t\,\dot t - \frac{1}{2}\,E\,(p_x^2 + 2\,m\,p_t), \nonumber \\
&&L_s(x,\,\dot x, \, t, \dot t) = \frac{{\dot x}^2}{2\,E} + \frac{m\,{\dot x}^2}{2\,\dot t}\Big[\frac{E\,m}{\dot t} - 1\Big],
\end{eqnarray}
where the trajectory of the free non-relativistic particle is embedded in a 2D configuration space characterized by the coordinates
$[x(\tau),\, t(\tau)]$ and the parameter $\tau$ specifies the trajectory of the particle as an evolution parameter. The momenta variables
$(p_x,\,p_t)$ are defined by: $p_x = ({\partial \,L}/{\partial\, \dot x})$, $p_t = ({\partial \,L}/{\partial\, \dot t})$ where 
$\dot x = ({d\,x}/{d\,\tau})$, $\dot t = ({d\,t}/{d\,\tau})$ are the generalized ``velocities" w.r.t. the coordinates $[x(\tau), t(\tau)]$
and $L$ stands for any of the {\it three} Lagrangians of Eq. (1). It is self-evident that the 4D phase space, corresponding to the 2D configuration space, 
is characterized by
$[x (\tau), t(\tau), p_x (\tau), p_t (\tau)]$. In the above equation (1), the mass of the non-relativistic particle is denoted by $m$
and $E(\tau)$ is the Lagrange multiplier variable that incorporates the constraint $p_x^2 + 2\,m\,p_t \approx 0$ in the Lagrangians $L_f$
and $L_s$. It is straightforward to note that $L_0$ and $L_s$ contain variables (and their first-order derivative) in the denominator {\it but} $L_f$ 
(i.e. the first-order Lagrangian) {\it does} not incorporate any variable (and/or its derivative) in {\it its} denominator. Furthermore, the starting 
Lagrangian\footnote {For a free massive NR particle, the standard action integral is: $S_t = \int_{- \infty}^{+ \infty} d\, t \, (\frac{1}{2}\,m \, {\dot x}^2)$
where $\dot x = (d x/dt)$ and time ``t" is the evolution parameter. This action has {\it no} reparameterization invariance.
If the evolution parameter is $ \tau $, then, the action integral is: $S_{\tau} = ({m}/{2})\,\int_{- \infty}^{+ \infty} d\, \tau \,
(d t/d \tau)\, (d x/d \tau)\,(d \tau / d t)\,(d x/d \tau)\,(d \tau / d t) $ which leads [32, 31] to the {\it final} action integral as: 
$S_{\tau} = (m / 2)\,\int_{- \infty}^{+ \infty} d\, \tau [(d x/d \tau)^2\,(d \tau/ d t )]  \, \equiv\, \int_{- \infty}^{+
 \infty} d\, \tau \, L_0 $. Hence, the starting Lagrangian becomes $ L_0(x,\,\dot x, \, t, \dot t) = \frac {m \, {\dot x}^2}{2\,\dot t}$ 
where $\dot x = (d x/ d \tau)$ and $\dot t = (d t/ d \tau)$.} $L_0$ does {\it not} permit the {\it massless}
$(m = 0)$ limit but the massless $(m = 0)$ limits are well-defined for $L_f$ and the second-order Lagrangian $L_s$. We would like to stress that the 
Lagrange multiplier variable behaves like the ``gauge" variable due to {\it its} transformation property.

For our further discussions, we shall concentrate on the first-order Lagrangian $L_f$ because it has maximum number of variables
(i.e. $x,\,p_x,\, t,\, p_t,\, E)$, allows massless limit and there are {\it no} variables (and/or their first-order derivative w.r.t. $\tau$) in {\it its} denominator. It is straightforward to 
check that under the following infinitesimal and continuous 1D diffeomorphism (i.e. reparameterization) symmetry transformations $(\delta_r)$, namely;
\begin{eqnarray}
&&\delta_r\,x = \epsilon\,\dot x, \quad\qquad \delta_r\,p_x = \epsilon\,{\dot p}_x, \quad\qquad \delta_r\,t = \epsilon\,\dot t, \quad\qquad \delta_r\,p_t 
= \epsilon\,{\dot p}_t,
\nonumber \\
&&\delta_r\,E = \frac{d}{d\,\tau}\,(\epsilon\,E), \quad\qquad \delta_r\,L_f = \frac{d}{d\,\tau}\,(\epsilon\,L_f),  
\end{eqnarray}
the action integral $S = \int_{- \infty}^{+ \infty} d\,\tau\,L_f$ remains invariant (i.e. $\delta_r\,S = 0$) for the physically well-defined variables in $L_f$
and the infinitesimal diffeomorphism transformation parameter $\epsilon(\tau)$ in: $\tau \longrightarrow \tau^{'} = g(\tau) = \tau - \epsilon(\tau)$
 where $g(\tau)$ is a {\it physically} well-defined function of $\tau$ such that it is {\it finite} at $\tau = 0$ and vanishes 
off at $\tau = \pm \infty$ . In fact, the infinitesimal and continuous reparameterization symmetry transformation $(\delta_r)$ is defined 
as: $\delta_r\,\phi(\tau) = \phi^{'}(\tau) - \phi(\tau)$ for the generic variable 
$\phi(\tau) = x(\tau),\, p_x(\tau),\, t(\tau),\, p_t(\tau),\, E(\tau)$ of our present theory.

The above infinitesimal and continuous reparameterization symmetry transformations (2) encompass in their folds the gauge symmetry transformations $(\delta_g)$
which are generated (see, e.g. [32] for details) by the first-class constraints: $\Pi_E \approx 0, \, (p_x^2 + 2\,m\,p_t) \approx 0$ where $\Pi_E$ is the canonical conjugate
momentum corresponding to the variable $E(\tau)$. Using the following Euler-Lagrange equations of motion (EL-EOMs) from $L_f$, namely;
\begin{eqnarray}
{\dot p}_x = 0, \qquad {\dot p}_t = 0, \qquad \dot x = E\,p_x, \qquad \dot t = E\,m,
\end{eqnarray}
{\it and} identifying the transformation parameters $\epsilon(\tau)\,E(\tau) = \xi(\tau)$, we obtain the infinitesimal and continuous {\it gauge}
symmetry transformations $(\delta_g)$, from the infinitesimal and continuous reparameterization symmetry transformations (2), as follows:
\begin{eqnarray}
\delta_g\,x = \xi\,p_x, \qquad \delta_g\,t = \xi\,m, \qquad \delta_g\,p_x = 0, \qquad \delta_g\,p_t = 0, \qquad \delta_g\,E = \dot\xi.
\end{eqnarray} 
It is elementary now to check that the first-order Lagrangian $(L_f)$ transforms to a total derivative under the infinitesimal and continuous gauge transformations $(\delta_g)$, namely;
\begin{eqnarray}
\delta_g\,L_f = \frac{d}{d\,\tau}\,\Big[\frac{\xi}{2}\,p_x^2\Big],
\end{eqnarray}
thereby rendering the action integral $S = \int^{+\infty}_{-\infty}d\,\tau\,L_f$ invariant (i.e $\delta_g \, S = 0$) under the 
infinitesimal and continuous gauge symmetry transformations $(\delta_g)$.

We end this subsection with the following decisive comments. First, the 1D diffeomorphism (i.e. reparameterization) transformations (2)
are {\it more} general than the infinitesimal and continuous {\it gauge} symmetry transformations (4). Second, the Lagrange multiplier variable
$E(\tau)$ behaves like a ``gauge" variable due to its transformation: $\delta_g\,E = \dot\xi$ in (4). Third, all the {\it three} Lagrangians
in (1) are {\it equivalent} and {\it all} of them respect the infinitesimal and continuous gauge and reparameterization symmetry
transformations [32]. Fourth, all the Lagrangians describe the free motion $({\dot p}_x = 0,\, {\dot p}_t = 0)$ of the NR particle. Hence, our
system is a non-relativistic free $( \dot p_x = \dot p_t = 0)$ particle. Fifth, the first-order Lagrangian $L_f$ is 
theoretically more interesting to handle because, as pointed out 
earlier, it incorporates the maximum number of variables. Finally, we can exploit the reparameterization {\it and} gauge symmetry 
transformations (2) and (4) for the BRST quantization. Following the usual BRST prescription, we note that the (anti-)BRST 
symmetry transformations, corresponding to the {\it classical} reparameterization symmetry transformation (2), are 
\begin{eqnarray}
s_{ab}\,x &=& \bar C\,\dot x,\quad s_{ab}\,p_x = \bar C\,\dot{p_x},\quad s_{ab}\,t = \bar C\,\dot t,\quad s_{ab}\,p_t = \bar C\; \dot p_t,\quad 
s_{ab}\,E = \frac{d}{d\,\tau}\,(\bar C\, E),\nonumber\\
s_{b}\,x &=& C\,\dot x,\quad~ s_{b}\,p_x = C\,\dot{p_x},\quad~ s_{b}\,t = C\,\dot t,\quad~~ s_{b}\,p_t = C\; \dot p_t,\quad 
s_{b}\,E = \frac{d}{d\,\tau}\,( C\, E),
\end{eqnarray}
where $(\bar C)C$ are the fermionic $(C^{2} = \bar C^{2} = 0,\, C\, \bar C + \bar C \, C = 0)$ (anti-)ghost variables 
corresponding to the {\it classical} infinitesimal diffeomorphism transformation parameter $\epsilon(\tau)$ [cf. Eq. (2)]. In exactly similar fashion, the 
(anti-)BRST symmetry transformations, corresponding to the {\it classical} gauge symmetry transformations (4), are
\begin{eqnarray}
s_{ab}\,x &=& \bar c\, p_x,\qquad s_{ab}\,p_x = 0,\qquad s_{ab}\,t = \bar c \, m,\qquad s_{ab}\,p_t = 0,\qquad s_{ab}\,E = \dot{\bar c},\nonumber\\
s_{b}\,x &=&  c\, p_x,\qquad s_{ab}\,p_x = 0,\qquad~ s_{b}\,t = c \, m,\qquad~ s_{b}\,p_t = 0,\qquad~ s_{b}\,E = \dot{c},
\end{eqnarray}
where the fermionic $(c^{2} = \bar c^{2} = 0,\, c\, \bar c + \bar c \, c = 0)$ variables $(\bar c)c$ are the (anti-)ghost variables 
corresponding to the {\it classical} gauge symmetry transformation parameter $\xi (\tau)$ of Eq. (4). In addition to the (anti-)BRST symmetry 
transformations in (6) and (7), we have the following {\it standard} (anti-)BRST transformations
\begin{eqnarray}
s_{ab}\,C &=& i\,\bar B, \qquad s_{ab}\,\bar B = 0, \qquad s_b\,\bar C = i\,B, \qquad s_b\,B = 0, \nonumber\\
s_{ab}\,c &=& i\,\bar b, \qquad~~ s_{ab}\,\bar b = 0, \qquad~~ s_b\,\bar c = i\,b, \qquad~~ s_b\,b = 0,  
\end{eqnarray} 
where the pairs $(B, \bar B)$ {\it and} $(b, \bar b)$ are the Nakanishi-Lautrup auxiliary variables in the contexts of the
BRST quantization of our reparameterization {\it and} gauge invariant system by exploiting the {\it classical} reparameterization and
gauge transformations, respectively.

\subsection{Quantum Nilpotent (Anti-)BRST Symmetries Corresponding to the Classical Gauge Symmetry Transformations}

We have listed the quantum (anti-)BRST symmetries corresponding to the {\it classical} gauge symmetry transformations (4)
in our Eqs. (7) and (8). It is elementary to check that these {\it quantum} symmetries are off-shell nilpotent
$(s_{(a)b}^2 = 0)$ of order two. The requirement of the absolute anticommutativity $(s_b\,s_{ab} + s_{ab}\,s_b = 0)$
leads to the restriction: $b + \bar b = 0 \Longrightarrow \bar b = -b$. As a consequence, we have the {\it full}
set of (anti-)BRST symmetry transformations [corresponding to the {\it classical} gauge symmetry transformations (4)] as follows:
\begin{eqnarray*}
&&s_{ab}\,x = \bar c\,p_x, \qquad s_{ab}\,p_x = 0, \qquad s_{ab}\,t = \bar c\,m, \qquad s_{ab}\,p_t = 0, \nonumber\\
&&s_{ab}\,E = \dot{\bar c}, \qquad s_{ab}\,\bar c = 0, \qquad s_{ab}\,c = - i\,b, \qquad s_{ab}\,b = 0, 
\end{eqnarray*} 
\begin{eqnarray}
&&s_{b}\,x = c\,p_x, \qquad s_{b}\,p_x = 0, \qquad s_{b}\,t =  c\,m, \qquad s_{b}\,p_t = 0, \nonumber\\
&&s_{b}\,E = \dot{c}, \qquad s_{b}\, c = 0, \qquad s_{b}\,\bar c =  i\,b, \qquad s_{b}\,b = 0. 
\end{eqnarray} 
It is straightforward to check that the above (anti-)BRST symmetry transformations are off-shell nilpotent $(s_{(a)b}^2 = 0)$
and absolutely anticommuting $(s_b\,s_{ab} + s_{ab}\,s_b = 0)$ in nature. The (anti-)BRST invariant Lagrangian 
$L_b$ (which is the generalization of the {\it classical} $L_f$ to its {\it quantum} level) can be written as\footnote {The structure of gauge-fixing and FP-ghost
terms is {\it exactly} like the Abelian 1-form $(A^{(1)} = dx^{\mu}\,A_\mu)$  gauge theory where we have the BRST-invariant Lagrangian density:
$ {\cal L}_{b} = -\,\frac{1}{4}\, F_{\mu \nu}\,F^{\mu \nu}
+ s_b\, [-\,i\, \bar c \,(\partial_{\mu}\,A^{\mu} + \frac{b}{2})]\, \equiv \, -\,\frac{1}{4}\, F_{\mu \nu}\,F^{\mu \nu}
+ s_{ab}\, [-\,i\, c \,(\partial_{\mu}\,A^{\mu} + \frac{b}{2})] \, \equiv \, -\,\frac{1}{4}\, F_{\mu \nu}\,F^{\mu \nu}
+ s_b \, s_{ab} \, [\frac{ i}{2} \, A^\mu A_\mu - \frac{1}{2} {\bar c\, c}]$. Here $A_\mu$ is the vector potential, $F_{\mu \, \nu} = \partial_{\mu}\,A_{\nu}
- \partial_{\nu}\, A_{\mu}$ is the field strength tensor and rest of the symbols are same
as in Eqs. (10) and (11). Note that the $2$-form $F^{(2)} = d \, A^{(1)} = \frac{1}{2}\,(d\,x^\mu \wedge d\, x^\nu )\,F_{\mu \nu}$ 
defines the field strength tensor $F_{\mu \nu}$ (where $d = d\,x^\mu \, \partial_{\mu}$ in $ F^{(2)} = d \, A^{(1)} $ stands for the exterior derivative of the differential geometry).}:
\begin{eqnarray}
L_b &=& L_f + s_b\,\Big[-\,i\,\bar c\,\Big(\dot E + \frac{b}{2}\Big)\Big] \equiv L_f + s_{ab}\,\Big[i\,c\,\Big(\dot E + \frac{b}{2}\Big)\Big], 
\nonumber\\
&\equiv&  L_f + s_b\,s_{ab}\,\Big[\frac{i\,E^2}{2} - \frac{\bar c\,c}{2}\Big] \equiv  L_f - s_{ab}\,s_{b}\,\Big[\frac{i\,E^2}{2} 
- \frac{\bar c\,c}{2}\Big]. 
\end{eqnarray}
In other words, we have expressed the gauge-fixing and Faddeev-Popov (FP) ghost terms in {\it three} different ways which,
ultimately, lead to the following expression for $L_b$, namely;
\begin{eqnarray}
L_b &=& L_f + b\,\dot E + \frac{b^2}{2} - i\,\dot{\bar c}\,\dot c, \nonumber\\
&\equiv& p_x\,\dot x + p_t\,\dot t - \frac{1}{2}\,E\,(p_x^2 + 2\,m\,p_t) + b\,\dot E + \frac{b^2}{2} - i\,\dot{\bar c}\,\dot c. 
\end{eqnarray}
It should be noted that we have dropped the total derivative terms in obtaining $L_b$ from (10). The above equation demonstrates that 
we have obtained a {\it unique} (anti-)BRST invariant Lagrangian. This has happened because
the CF-type restriction is trivial (i.e. $b + \bar b = 0$) in our {\it simple} case of NR system. We can explicitly check that:
\begin{eqnarray}
s_b\,L_b = \frac{d}{d\,\tau}\,\Big[\frac{c}{2}\,p_x^2 + b\,\dot c\,\Big], \qquad s_{ab}\,L_b = \frac{d}{d\,\tau}\,
\Big[\frac{\bar c}{2}\,p_x^2 + b\,\dot{\bar c}\,\Big], 
\end{eqnarray}
which lead to the derivation of the conserved (anti-)BRST charges $[Q_{(a)b}]$ as follows:
\begin{eqnarray}
Q_{ab} = b\,\dot{\bar c} + \frac{1}{2}\,\bar c \,(p_x^2 + 2\,m\,p_t) \equiv b\,\dot{\bar c} - \dot b\,\bar c, \nonumber \\
Q_{b} = b\,\dot{c} + \frac{1}{2}\,c \,(p_x^2 + 2\,m\,p_t) \equiv b\,\dot{c} - \dot b\,c.
\end{eqnarray}
In the last step, we have used $\dot b = -\,(1/2)\,(p_x^2 + 2\,m\,p_t)$ which emerges out as the EL-EOM from $L_b$ w.r.t. the 
Lagrange multiplier variable $E(\tau)$ .

We close this sub-section with a few crucial and decisive remarks. First, we can check that the (anti-)BRST charges are
conserved $[{\dot Q}_{(a)b} = 0]$ by using the EL-EOMs. Second, the (anti-)BRST charges $[Q_{(a)b}]$ are off-shell
nilpotent $[Q_{(a)b}^2 = 0]$ of order {\it two} due to the {\it direct} observations that: $s_b\,Q_b = s_b\,[b\,\dot c - \dot b\,c] = 0$
and $s_{ab}\,Q_{ab} = s_{ab}\,[b\,\dot{\bar c} - \dot b\,\bar c] = 0$ which encode in their folds $s_b\,Q_b = -\,i\,\{Q_b, Q_b\}
 = 0 \Longrightarrow Q_b^2 = 0$ {\it and} $s_{ab}\,Q_{ab} = -\,i\,\{Q_{ab}, Q_{ab}\} = 0 \Longrightarrow Q_{ab}^2 = 0$.
 Third, the above nilpotency is {\it also} encoded in: $Q_b = s_b\,[b\,E + i\,\dot{\bar c}\,c]$ implying that $s_b\,Q_b = 0$
due to $s_b^2 = 0$ {\it and} we {\it also} point out that $s_{ab}\,Q_{ab} = 0$ due to the nilpotency $(s_{ab}^2 = 0)$ of $s_{ab}$ because
$Q_{ab} = s_{ab}\,[b\,E + i\,\bar c \,\dot c]$. Fourth, we observe that $s_{ab}\,Q_b = i\,\{Q_b, Q_{ab}\} \equiv -\,i\,b\,
\dot b + i\,\dot b\,b = 0$ and $s_b\,Q_{ab} = -\,i\,\{Q_{ab}, Q_b\} = i\,b\,\dot b - i\,b\,\dot b = 0$ which explicitly
lead to the conclusion that the off-shell nilpotent charges $Q_{(a)b}$ are {\it also} absolutely anticommuting $(Q_b\,Q_{ab} + Q_{ab}\,Q_b
 = 0)$ in nature. Fifth, the above observation of the absolute anticommutativity can be also expressed in terms of 
the nilpotency property because we observe that $Q_b = s_{ab}\,(-\,i\,\dot c\,c)$ and $Q_{ab} = s_b\,(i\,\dot{\bar c}\,\bar c)$
which imply that $s_{ab}\,Q_b = -\,i\,\{Q_b, Q_{ab}\} = 0$ and $s_{b}\,Q_{ab} = -\,i\,\{Q_{ab}, Q_{b}\} = 0$ [due to the
off-shell nilpotency $(s_{ab}^2 = 0)$ of the anti-BRST as well as the off-shell nilpotency $(s_b^2 = 0)$ of the BRST symmetry transformations].
Sixth, it can be seen that the physical space (i.e. $\mid phys>$) in the {\it total} Hilbert space of states is defined by
$Q_b \mid phys> = 0$ which implies that $b\mid phys> \equiv \Pi_{E}\, \mid phys> = 0$ and $\dot b \mid phys> \equiv (p_x^2 + 2\,m\,p_t)\,\mid phys> = 0$.
In other words, the Dirac quantization conditions (with the first-class constraints $\Pi_{E} \approx 0, \, p_x^2 + 2\,m\,p_t \approx 0$) are 
beautifully satisfied. Finally, physicality criterion $Q_b \mid phys> = 0$ implies that the two physical states $\mid phys^{'}>$ and $\mid phys>$
belong to the {\it same} cohomological class 
w.r.t. the nilpotent BRST charge $Q_b$ if they differ by a BRST {\it exact} state (i.e. $\mid phys^{'}> = \mid phys> + Q_b\,\mid \chi>$
for non-null $|\chi>$).

\section{Nilpotent and Anticommuting (Anti-)BRST Symmetries for the Phase Variables: MBTSA}

This section is devoted to the derivation of the transformations: $s_b\,x = C\,\dot x,\, s_b\,p_x = C\,{\dot p}_x, \, s_b\,t = C\,\dot t,
\, s_b\,p_t = C\,{\dot p}_t,\, s_{ab}\,x = \bar C\,\dot x,\, s_{ab}\,p_x = \bar C\,{\dot p}_x, \, s_{ab}\,t = \bar C\,\dot t,
\, s_{ab}\,p_t = \bar C\,{\dot p}_t$ by exploiting the theoretical tricks of MBTSA. Before we set out to perform this exercise, it is essential to pinpoint 
the off-shell nilpotency and absolute anticommutativity properties of the (anti-)BRST symmetry transformations on the phase variables [cf. Eq. (6)].
It can be easily checked that the off-shell nilpotency requirement (i.e. $s_{(a)b}^{2}\,S = 0,\, S = x,\, p_x,\, t,\,p_t $) leads to the (anti-)BRST
symmetry transformations for the (anti-)ghost variables as:  
\begin{eqnarray}
s_{ab}\,\bar C = \bar C \,\dot {\bar C},  \qquad\qquad  s_b\,C = C\,\dot C.
\end{eqnarray}
Furthermore, the absolute anticommutativity requirement: $ \{s_b, \,s_{ab} \}S = 0$ for the generic phase variable $S = x,\, p_x,\, t,\,p_t$ 
leads to the following
\begin{eqnarray}
\{s_b,\, s_{ab}\}S = i\,[B + \bar B + i\,(\bar C\, \dot C - \dot {\bar C}\, C)]\,\dot S \,\quad \Longrightarrow \quad \, B + 
\bar B + i\,(\bar C\, \dot C - \dot {\bar C}\, C) = 0.
\end{eqnarray}
In other words, the absolute anticommutativity property ($s_b\,s_{ab} + s_{ab}\,s_b = 0$) is satisfied if and only if we invoke the
sanctity of the CF-type restriction: $ B + \bar B + i\,(\bar C\, \dot C - \dot {\bar C}\, C) = 0$. It goes without saying that the {\it above} cited
requirements of the off-shell
nilpotency and  absolute anticommutativity properties are very {\it sacrosanct} within the framework of BRST approach to gauge and/or reparameterization
invariant theories.

Against the backdrop of the above discussions, we set out to deduce the (anti-)BRST symmetry transformations: $s_{ab}\,S = \bar C\,\dot S, \, s_b\,S
= C\, \dot S$ (with $S = x,\, p_x,\, t,\,p_t$) and the CF-type restrictions:  $ B + \bar B + i\,(\bar C\, \dot C - \dot {\bar C}\, C) = 0$
within the framework of MBTSA. Towards this end in our mind, first of all, we generalize the 
{\it classical} function $g(\tau)$ [in $\tau \longrightarrow \tau^{'} = g(\tau) 
\equiv \tau - \epsilon(\tau)$] onto a $(1, 2)$-dimensional supermanifold as
\begin{eqnarray}
g(\tau) \quad \longrightarrow \quad \tilde g (\tau, \theta, \bar\theta) = \tau - \theta\,\bar C(\tau) - \bar\theta\,C(\tau) + \theta\,\bar\theta\,k(\tau),
\end{eqnarray} 
where $(\bar C)C$ variables are the (anti-)ghost variables of Eq. (6) and $k(\tau)$ is a secondary variable that has to be determined from 
the consistency conditions [that include the off-shell nilpotency as well as absolute anticommutativity requirements]. It will be noted that,
due to the mappings: $s_b \leftrightarrow \partial_{\bar\theta}\mid_{\theta = 0},\, s_{ab} \leftrightarrow \partial_{\theta}\mid_{\bar\theta = 0}$ [14-16],
we have taken the coefficients of $\theta$ and $\bar\theta$ in Eq. (16) as the (anti-)ghost variables $(\bar C)C$. This has been done due to our
observation in the infinitesimal reparameterization symmetry transformation $(\delta_r)$ [where $\delta_r\,\tau = -\,\epsilon(\tau)$] at the {\it classical} 
level. Following the basic tenet of BRST formalism, the infinitesimal parameter $\epsilon(\tau)$
has been replaced (in the BRST-{\it quantized} theory) by the (anti-)ghost variables thereby leading to the (anti-)BRST symmetry transformations: 
$s_{ab}\,\tau = -\,\bar C,\,s_b\,\tau = -\,C$.

For our present 1D diffeomorphism (i.e. reparameterization) invariant theory, the generic variable $S(\tau)$ can be generalized to
a supervariable [$\tilde S(\tilde g(\tau, \theta, \bar\theta), \theta, \bar\theta)$] on the (1, 2)-dimensional supermanifold [22] with the following 
super expansion along {\it all} the Grassmannian directions of the (1, 2)-dimensional supermanifold, namely;
\begin{eqnarray} 
\tilde S \big[\tilde g(\tau, \theta, \bar\theta), \theta, \bar\theta) \big] = {\cal S} \big [\tilde g(\tau, \theta, \bar\theta) \big] 
+ \theta\,\bar R \big[\tilde g(\tau, \theta, \bar\theta) \big] 
+ \bar\theta\,R \big [\tilde g(\tau, \theta, \bar\theta) \big] 
+ \theta\,\bar\theta\,Q \big [\tilde g(\tau, \theta, \bar\theta) \big],
\end{eqnarray}
where the expression for $\tilde g(\tau, \theta, \bar\theta)$ is given in Eq. (16). It should be noted that {\it all} the primary as well as the secondary supervariables on the
r.h.s. of (17) are function of the $(1, 2)$-dimensional super infinitesimal diffeomorphism transformation (16). At this stage, we can perform the Taylor
expansions for {\it all} the supervariables as:
\begin{eqnarray}
\theta\,\bar\theta\,Q(\tau - \theta\,\bar C - \bar\theta\,C + \theta\,\bar\theta\,k) &=& \theta\,\bar\theta\,Q(\tau), \nonumber\\
\bar\theta\,R(\tau - \theta\,\bar C - \bar\theta\,C + \theta\,\bar\theta\,k) &=& \bar\theta\,R(\tau) + \theta\,\bar\theta \, \bar C(\tau)\,\dot R(\tau),
 \nonumber\\
\theta\,\bar R(\tau - \theta\,\bar C - \bar\theta\,C + \theta\,\bar\theta\,k) &=& \theta\,\bar R(\tau) - \theta\,\bar\theta \,  C(\tau)\,\dot{\bar R}(\tau),
 \nonumber\\
{\cal S} (\tau - \theta\,\bar C - \bar\theta\,C + \theta\,\bar\theta\,k) &=& S(\tau) - \theta\,\bar C(\tau)\,\dot S(\tau) - \bar\theta\,C(\tau)\,\dot S(\tau)
\nonumber\\ 
&+& \theta\,\bar\theta\, \big [k(\tau)\,\dot S (\tau) - \bar C (\tau) \, C (\tau)\,\ddot S (\tau) \big ].
\end{eqnarray} 
Collecting all these terms and substituting them into (17), we obtain the following super expansion for the supervariable on the (1, 2)-dimensional supermanifold, namely;
\begin{eqnarray}
\tilde S \big [\tilde g(\tau, \theta, \bar\theta), \theta, \bar\theta \big ] &=& S(\tau) + \theta\,(\bar R - \bar C\,\dot S) + \bar\theta\,(R - C\,\dot S) \nonumber\\
&+& \theta\,\bar\theta\,\big[Q + \bar C\,\dot R - C\,\dot{\bar R} + k\,\dot S + C\,\bar C\,\ddot S\big].
\end{eqnarray}
We now exploit the horizontality condition (HC) which physically implies that {\it all} the {\it scalar} variables should {\it not} transform { \it at all} under
any kind of spacetime, internal, supersymmetric, etc., transformations. With respect to the 
1D space of trajectory of the particle, all the supervariables on
the l.h.s. and r.h.s. of Eq. (18) are {\it scalars}. The HC, in our case, is:
\begin{eqnarray}
\tilde S \big [\tilde g(\tau, \theta, \bar\theta), \theta, \bar\theta) \big]  = S(\tau).
\end{eqnarray}
The above equality implies that {\it all} the coefficients of $\theta$, $\bar \theta$ and $\theta\, \bar \theta$ of Eq. (19) should be set equal to zero.
In other words, we have the following:
\begin{eqnarray}
R = C \, \dot S, \qquad \bar R =\bar C \, \dot S, \qquad Q = C \, \dot {\bar R} - \bar C \, \dot R - k \, \dot S + \bar C \, C \, \ddot S.
\end{eqnarray}
Substitutions of the values of $R$ and $\bar R$ into the expression for $Q$ leads to the following:
\begin{eqnarray}
Q = -\,(\bar C \, \dot C + \dot {\bar C} \, C)\, \dot S - k \, \dot S - \bar C \, C \, \ddot S.
\end{eqnarray}
As explained before Eq. (20) (i.e. exploiting the key properties of {\it scalars}), it is evident that (17) can be {\it finally} written 
(with $\tilde S \big [\tilde g(\tau, \theta, \bar\theta), \theta, \bar\theta \big ] = \tilde S (\tau, \theta, \bar \theta) $ as
\begin{eqnarray}
\tilde S\big[ \tau, \theta, \bar\theta)\big] &=& {S}(\tau) + \theta\,\bar R(\tau) 
+ \bar\theta\,R(\tau) + \theta\,\bar\theta\,Q(\tau) \nonumber\\
& \equiv & {S}(\tau) + \theta \,(s_{ab}\,S) + \bar \theta \,(s_{b}\,S) + \theta \,\bar \theta \,(s_{b} \, s_{ab} \, S),
\end{eqnarray}
where, due to the well known mappings: $s_b\leftrightarrow \partial_{\bar \theta}\mid _{\theta = 0},\, s_{ab}\leftrightarrow 
\partial_{\theta}\mid _{\bar \theta = 0}$ [14-16],
the coefficients of $\theta$ and $\bar \theta$ are the anti-BRST and BRST symmetry transformations [cf. Eq. (6)]. We point out that the key properties
of {\it scalars} on the r.h.s. of Eq. (17) implies that we have: ${\cal S}[\tilde g(\tau, \theta, \bar\theta)] = S(\tau), \, R[\tilde g(\tau, 
\theta, \bar\theta)] = R(\tau), \, \bar R[\tilde g(\tau, \theta, \bar\theta)] = \bar R(\tau)$ and $Q[\tilde g(\tau, \theta, \bar\theta)] = Q(\tau)$.

A comparison between (21) and (23) implies that we have already derived the nilpotent (anti-)BRST symmetry transformations: $R = s_b\,S = C\,\dot S$ and
$\bar R = s_{ab}\,S = \bar C\,\dot S$. In other words, we have obtained: $s_b\,x = C\,\dot x, \, s_b\,p_x = C\,{\dot p}_x, \, s_b\,t = C\,\dot t, \,
s_b\,p_t = C\,{\dot p}_t$ and $s_{ab}\,x = \bar C\,\dot x, \, s_{ab}\,p_x = \bar C\,{\dot p}_x, \, s_{ab}\,t = \bar C\,\dot t, \,
s_{ab}\,p_t = \bar C\,{\dot p}_t$. Furthermore, it is evident that:
\begin{eqnarray}
s_b\,s_{ab}\,S = Q = -\,(\bar C\,\dot C + \dot{\bar C}\,C)\,\dot S - k\,\dot S - \bar C\,C\,\ddot S.
\end{eqnarray}
The requirement of the absolute anticommutativity: $\{s_b, s_{ab}\}\,S = 0$ implies that $s_b\,s_{ab}\,S = -\,s_{ab}\,s_b\,S$ which, in turn,
leads to the following relationships:
\begin{eqnarray}
s_b\,s_{ab}\,S &=& s_b\,\bar R = Q \equiv  -\,(\bar C\,\dot C + \dot{\bar C}\,C)\,\dot S - k\,\dot S - \bar C\,C\,\ddot S, \nonumber\\
-s_{ab}\,s_b\,S &=& -\,s_{ab}\,R = Q \equiv -\,(\bar C\,\dot C + \dot{\bar C}\,C)\,\dot S - k\,\dot S - \bar C\,C\,\ddot S. 
\end{eqnarray}
The explicit computation of the following, using the (anti-)BRST symmetry transformations of the phase variables in Eqs. (6) and (14), are:
\begin{eqnarray}
s_b\,\bar R &=& i\,B\,\dot S - \bar C\,\dot C\,\dot S - \bar C\,C\,\ddot S \equiv Q, \nonumber\\
-\,s_{ab}\, R &=& -\,i\,\bar B\,\dot S - \dot{\bar C}\,C\,\dot S - \bar C\,C\,\ddot S \equiv Q.
\end{eqnarray}
Equating (25) and (26), we obtain the following interesting relationship:
\begin{eqnarray}
k = -\,\dot{\bar C}\,C - i\,B \equiv i\,\bar B - \bar C\,\dot C \quad \Longrightarrow \quad B + \bar B + i\,(\bar C\,\dot C - \dot{\bar C}\,C) = 0.
\end{eqnarray}
In other words, it is the consistency conditions of the BRST formalism that lead to the determination of $k(\tau)$ in Eq. (16) within the ambit of MBTSA. A
close look at Eqs. (25), (26) and (27) establishes that a precise determination of $Q(\tau)$ in (23) leads to (i) the validity 
of the absolute anticommutativity (i.e. $\{s_b, s_{ab}\}\,S = 0$) of the off-shell nilpotent (anti-)BRST symmetries, and (ii) the deduction of the 
(anti-)BRST invariant\footnote{This statement is {\it true} only when the whole theory is considered on a {\it submanifold} of the
Hilbert space of the quantum variables where the CF-type restriction: $B + \bar B + i\,(\bar C\,\dot C - \dot{\bar C}\,C) = 0$ is satisfied. In other words, 
we explicitly compute $ s_b \, [B + \bar B + i\,(\bar C\,\dot C - \dot{\bar C}\,C)] = (\frac{d}{d \tau}) \, [B + \bar B + i\,(\bar C\,\dot C - 
\dot{\bar C}\,C)]\,C - [B + \bar B + i\,(\bar C\,\dot C - \dot{\bar C}\,C)]\, \dot C $ and $ s_{ab} \, [B + \bar B + i\,(\bar C\,\dot C -
\dot{\bar C}\,C)] = (\frac{d}{d \tau}) \, [B + \bar B + i\,(\bar C\,\dot C - 
\dot{\bar C}\,C)]\, \bar C - [B + \bar B + i\,(\bar C\,\dot C - \dot{\bar C}\,C)]\, \dot{\bar C} $ which imply that $ s_{(a)b}\,
[B + \bar B + i\,(\bar C\,\dot C - \dot{\bar C}\,C)] = 0$ is true {\it only} on the above mentioned submanifold.} CF-type restriction 
$B + \bar B + i\,(\bar C\,\dot C - \dot{\bar C}\,C) = 0$ on our theory.

We conclude this section with the following useful and crucial remarks. First, we set out to derive the (anti-)BRST symmetry transformations
(corresponding to the classical reparameterization symmetry transformations) for the phase variables [cf. Eq. (6)]. We have accomplished
this goal in Eq. (21). Second, we have derived the CF-type restriction: $B + \bar B + i\,(\bar C\,\dot C - \dot{\bar C}\,C) = 0$ within the
purview of MBTSA [cf. Eq. (27)] which is actually hidden in the determination of $Q(\tau)$ in Eq. (23). Third, for the application of the theoretical
potential of MBTSA, we have taken the {\it full} super expansion of the {\it generic} supervariable [cf. Eq. (17)] along {\it all} the possible
Grassmannian directions of the (1, 2)-dimensional supermanifold. Fourth, unlike the application of BT-superfield/supervariable
approach to the {\it gauge} theories [14-16] where spacetime does {\it not} change, in the case of MBTSA, the super diffeomorphism transformation
(16) has been taken into account in all the {\it basic} as well as {\it secondary} supervariables. Fifth, taking into account 
the inputs from Eqs. (21) and (26), we obtain the following super expansion of the {\it generic} variable $S(\tau)$, namely;
\begin{eqnarray}
\tilde S^{(h)} (\tau, \theta, \bar\theta) &=& S(\tau) + \theta\,(\bar C\,\dot S) + \bar\theta\,(C\,\dot S) + \theta\,\bar\theta\,[i\,B\,\dot S - \bar C\,\dot C
\,\dot S - \bar C\,C\,\ddot S] \nonumber\\
&\equiv & S(\tau) + \theta\,(s_{ab}\,S) + \bar\theta\,(s_b\,S) + \theta\,\bar\theta\,(s_b\,s_{ab}\,S),
\end{eqnarray}
where $S = x, p_x, t, p_t$ and the superscript $(h)$ on the supervariable $\tilde S(\tau, \theta, \bar\theta)$ denotes that this supervariable 
has been obtained after the application of HC. Finally, the standard nilpotent (anti-)BRST symmetry transformations
(8) dictate that we can have the following (anti-)chiral super expansions for the supervariables corresponding to $(\bar C)C$, namely;
\begin{eqnarray}
&&C(\tau) \quad \longrightarrow \quad {F}^{(c)}(\tau, \theta) = C (\tau) + \theta\,(i\,\bar B) \equiv C (\tau) + \theta\,(s_{ab}\,C), \nonumber\\
&&\bar C(\tau) \quad \longrightarrow \quad {{\bar F}^{(ac)}}(\tau,  \bar \theta) = \bar C (\tau) + \bar\theta\,(i\,B) \equiv \bar C (\tau) 
+ \bar\theta\,(s_{b}\,\bar C), 
\end{eqnarray}   
where the superscripts $(c)$ and $(ac)$ denote the {\it chiral} and {\it anti-chiral} supervariables. The above observation gives us a clue that we should exploit
the theoretical strength of ACSA to BRST formalism for our further discussions.\\

\section{Coupled Lagrangians and Quantum (Anti-)BRST Symmetries Corresponding to the Classical Reparameterization Symmetry Transformations}

In addition to the quantum (anti-)BRST symmetries in (6), (8) and (14), we derive {\it all} the other off-shell nilpotent and 
absolutely anticommuting (anti-)BRST symmetries corresponding to the {\it classical} infinitesimal and continuous reparameterization
symmetry transformations (2). We exploit the strength of the {\it sacrosanct} requirements of off-shell nilpotency and absolute anticommutativity
properties. In this context, we point out that we have already derived $s_b\,C = C\,\dot C, \, s_{ab}\,\bar C = \bar C\,\dot{\bar C}$ 
by invoking the sanctity of the off-shell nilpotency $(s_{(a)b}^2 = 0)$ property for the phase variables (i.e. $s_{(a)b}^2\,S = 0, 
S = x, p_x, t, p_t$). It is interesting to note the following absolute anticommutativity requirements, namely;
\begin{eqnarray}
&&\{s_b, s_{ab}\}\,C = 0 \quad \Longrightarrow \quad s_b\,\bar B = \dot{\bar B}\,C - \bar B\,\dot C, \nonumber \\
&&\{s_b, s_{ab}\}\,\bar C = 0 \quad \Longrightarrow \quad s_{ab}\,B = \dot{B}\,\bar C - B\,\dot{\bar C}, 
\end{eqnarray}
leads to the derivation of the $s_{b}\,{\bar B}$ and $s_{ab}\,B$. We can readily check that $s_b^2\,\bar B = 0, \, s_{ab}^2\,B = 0$ are satisfied due to our knowledge of the BRST and anti-BRST symmetry transformations:
 $s_b\,C = C\,\dot C, \, 
s_{ab}\,\bar C = \bar C\,\dot{\bar C}$ {\it and} the fermionic $(C^2 = {\bar C}^2 = 0, \, C\,\bar C + \bar C\,C = 0)$ nature 
of the (anti-)ghost variables $(\bar C)\,C$. We further note that $\{s_b, s_{ab}\}\,B = 0$ and $\{s_b, s_{ab}\}\,\bar B = 0$.
The requirement of the absolute anticommutativity on the $E(\tau)$ variable leads to:
\begin{eqnarray}
\{s_b, s_{ab}\}\,E(\tau) = \frac{d}{d\,\tau}\,\Big[i\,\big\{B + \bar B + i\,(\bar C\,\dot C - \dot{\bar C}\,C)\big\}\,E(\tau)\Big].
\end{eqnarray}  
Thus, we emphasize that the absolute anticommutativity property $(s_b\,s_{ab} + s_{ab}\,s_b = 0)$ on the phase variables [cf. Eq. (15)]
as well as on the Lagrange multiplier variable [cf. Eq. (31)] are satisfied if and only if the CF-type restriction is invoked. In the full blaze of glory,
the {\it quantum} (anti-)BRST symmetry transformations [corresponding to the infinitesimal reparameterization symmetry transformations (2)] are as follows:     
\begin{eqnarray}
s_{ab}\,x &=& \bar C\,\dot x,\quad s_{ab}\,p_x = \bar C\,\dot{p_x},\quad s_{ab}\,t = \bar C\,\dot t,\quad s_{ab}\,p_t = \bar C\; \dot p_t,\quad 
s_{ab}\,E = \frac{d}{d\,\tau}\,(\bar C\, E),\nonumber\\
s_{ab}\,C &=& i\, \bar B,\quad s_{ab}\, \bar C = \bar C\, \dot {\bar C}, \quad s_{ab}\,\bar B = 0, \quad  s_{ab}\, B = \dot {B}\, \bar C - B\, \dot{\bar C},
\end{eqnarray}
\begin{eqnarray}
s_{b}\,x &=& C\,\dot x,\quad~ s_{b}\,p_x = C\,\dot{p_x},\quad~ s_{b}\,t = C\,\dot t,\quad~~ s_{b}\,p_t = C\; \dot p_t,\quad 
s_{b}\,E = \frac{d}{d\,\tau}\,( C\, E),\nonumber\\
s_{b}\,\bar C &=& i\, B,\quad s_{b}\, C = C\, \dot C, \quad s_{b}\, B = 0, \quad  s_{b}\, \bar B = \dot {\bar B}\, C - \bar B\, \dot C.
\end{eqnarray}
The above { \it fermionic} symmetry transformations are off-shell nilpotent and absolutely anticommuting provided the whole theory is considered on
a submanifold of the space of quantum variables where the CF-type restriction: $B + \bar B + i\, (\bar C\, \dot C - \dot {\bar C}\, C) = 0$ is satisfied.

The existence of the {\it above} CF-type restriction leads to the derivation of the coupled (but equivalent) Lagrangians (i.e. $L_B$ and $L_{\bar B}$) as follows:
\begin{eqnarray}
L_B = L_f + s_b\,s_{ab}\Big[\frac{i\,E^2}{2} - \frac{\bar C\, C}{2}\Big],\nonumber\\
L_{\bar B} = L_f - s_{ab}\,s_{b}\Big[\frac{i\,E^2}{2} - \frac{\bar C\, C}{2}\Big].
\end{eqnarray}
We point out that the terms inside the square brackets are {\it same} as in Eq. (10) for the BRST analysis of the {\it classical} 
gauge symmetry transformations (4). Furthermore, in contrast to the {\it unique} (anti-)BRST invariant Lagrangian [cf. Eq. (11)] (corresponding
to the {\it classical} gauge symmetry transformations), we have obtained here a set of coupled (but equivalent) (anti-)BRST invariant
Lagrangians in Eq. (34). This has happened because of the fact that the CF-type restriction ($b + \bar b = 0 $) is {\it trivial}
in the case of the {\it former} while it is a {\it non-trivial} restriction [$B + \bar B + i\, (\bar C \, \dot C - \dot {\bar C} \, C) = 0 $] in
the context of the {\it latter}.

One can readily compute the operation of $s_{(a)b}$ on the quantities in the square brackets of Eq. (34). In the full blaze of their glory, 
the coupled (but equivalent) Lagrangians $L_B$ and $L_{\bar B}$ are as follows\footnote{ It will be worthwhile to mention {\it here} that the {\it form} of the gauge-fixing and Faddeev-Popov ghost terms is {\it same} as in the cases of NSUSY (i.e. scalar) and SUSY (i.e. spinning) relativistic particles [23, 24].}
\begin{eqnarray}
L_B &=& L_f  + B\,\Big[E\,\dot E -i\, (2\, \dot{\bar C}\, C  + {\bar C}\,\dot C)\Big]+ \frac{B^2}{2} \nonumber\\
&&- i\,E\,\dot E\,\dot{\bar C}\,C - \,i\,E^2\,\dot{\bar C}\,\dot C  - \dot{\bar C}\,{\bar C}\,\dot C\,C, \nonumber\\
L_{\bar B} &=& L_f  - \bar B\,\Big[E\,\dot E - i\,(2\, {\bar C}\,\dot C +  \dot{\bar C}\,C)\Big]+\frac{{\bar B}^2}{2} \nonumber\\
&&- i\,E\,\dot E\,{\bar C}\,\dot C - \,i\,E^2\,\dot{\bar C}\,\dot C  - \dot{\bar C}\,{\bar C}\,\dot C\,C,
\end{eqnarray}
where the subscripts $B$ and $\bar B$ on the Lagrangians are appropriate because $L_B$ depends {\it uniquely} on the Nakanishi-Lautrup 
auxiliary variable $B$ (where ${\bar B}$ is {\it not} present at all). Similarly, the Lagrangian $L_{\bar B}$ is {\it uniquely} dependent on $\bar B$.
They are coupled because the EL-EOMs with respect to $B$ and $\bar B$ from $L_B$ and $L_{\bar B}$, respectively, yield
\begin{eqnarray}
B = -\,E\,\dot E +2\,i\,\dot{\bar C}\,C + i\,\bar C\,\dot C, \qquad \bar B = E\,\dot E - 2\,i\,{\bar C}\,\dot C - i\,\dot{\bar C}\, C, 
\end{eqnarray}
which lead to the deduction of the CF-type restrictions: $B + \bar B + i \,(\bar C \, \dot C - \dot {\bar C} \, C) = 0 $. Furthermore, the condition 
$L_B \equiv L_{\bar B}$ also demonstrates the existence of the CF-type restriction: $B + \bar B + i \,(\bar C \, \dot C - \dot {\bar C} \, C) = 0 $ 
on our theory (cf. Appendix A below).

At this stage, we are in the position to study the (anti-)BRST symmetries of the Lagrangians $L_B$ and $L_{\bar B}$. It is straightforward
to note that we have the following:
\begin{eqnarray}
&&s_b\,L_B = \frac{d}{d\,\tau}\Big[C\,L_f + B^2\,C - i\,B\,\bar C\,\dot C\,C + E\,\dot E\,B\,C  + E^2\,B\,\dot C \Big],
\end{eqnarray}
\begin{eqnarray}
&&s_{ab}\,L_{\bar B} = \frac{d}{d\,\tau}\Big[\bar C\,L_f + {\bar B}^2\,\bar C  
- i\,\bar B\,\dot{\bar C}\,\bar C\,C - E\,\dot E\,\bar B\,\bar C - E^2\,\bar B\,\dot{\bar C}  \Big].
\end{eqnarray}
The above observations demonstrate that the action integrals $S_1 = \int_{-\infty}^{\infty} d\,\tau\,L_B$ and $S_2 = \int_{-\infty}^{\infty} d\,\tau\,
L_{\bar B}$ remain invariant under the SUSY-type (i.e. fermionic) off-shell nilpotent, continuous and infinitesimal (anti-)BRST symmetry transformations for
the physical variables that vanish off at $\tau = \pm \infty$.
At this crucial juncture, we establish the {\it equivalence} of the coupled Lagrangian $L_B$ and $L_{\bar B}$ w.r.t the (anti-)BRST symmetry transformations
$[s_{(a)b}]$. In this context, we apply $s_{ab}$ on $L_B$ and $s_{b}$ on $L_{\bar B}$ to obtain the following 
\begin{eqnarray}
s_{ab}\,L_{B} &=& \frac{d}{d\,\tau}\,\Big[\bar C\,L_f  
+ E\,\dot E\,(i\,\dot{\bar C}\,\bar C\,C + B\,\bar C) + E^2\,(i\,\dot{\bar C}\,\bar C\,\dot C + B \,\dot{\bar C}) \nonumber\\ 
&+&  {B}^2\,\bar C + i\,(2\,B - \bar B)\,\dot{\bar C}\,\bar C\,C \Big] \nonumber\\
&+& \big[B+ \bar B + i\,(\bar C\,\dot C - \dot{\bar C}\,C)\big]\,( 2\,i\,\dot{\bar C}\,\bar C\,\dot C
 - 2\,B\,\dot{\bar C}- E\,\dot E\,\dot{\bar C} + i\,\ddot{\bar C}\,\bar C\,C ) \nonumber\\
&-& \frac{d}{d\,\tau}\big[B+ \bar B + i\,(\bar C\,\dot C - \dot{\bar C}\,C)\big]\, \big[ B\,\bar C + E^2\,\dot{\bar C} \big],
\end{eqnarray}
\begin{eqnarray}
s_b\,L_{\bar B} &=& \frac{d}{d\,\tau}\,\Big[C\,L_f + E\,\dot E\,(i\,\bar C\,C
\,\dot C - \bar B\,C) + E^2\,(i\,\dot{\bar C}\,C\,\dot C - \bar B \,\dot C)  \nonumber\\ 
&+& {\bar B}^2\,C - i\,(2\,\bar B - B)\,\bar C\,C\,\dot C \Big] \nonumber\\
&+& \big[B+ \bar B + i\,(\bar C\,\dot C - \dot{\bar C}\,C)\big]\,( - 2\,i\,
\dot{\bar C}\,C\,\dot C - 2\,\bar B\,\dot C + E\,\dot E\,\dot C + i\,\bar C\,\ddot C\,C ) \nonumber\\
&+& \frac{d}{d\,\tau}\big[B+ \bar B + i\,(\bar C\,\dot C - \dot{\bar C}\,C)\big]\, \big[+ E^2\,\dot C - \bar B\,C  \big],
\end{eqnarray}
which demonstrate that the coupled Lagrangians $L_B$ and $L_{\bar B}$ (and corresponding action integrals) respect {\it both} 
(i.e. BRST and anti-BRST) symmetry transformations {\it together} provided the whole theory is considered on a supermanifold in the Hilbert space
 of {\it quantum} variables where the CF-type restriction: $B + \bar B + i \, (\bar C \, \dot C - \dot {\bar C} \, C) = 0 $ is satisfied. It should 
be recalled that, under the {\it latter} restriction, we {\it also} have the absolute anticommutativity property (i.e. $\{s_b, s_{ab}\} = 0$ of the (anti-)BRST
symmetry transformations.

We end this section with the following key comments. First, the properties of the off-shell nilpotency and absolute anticommutativity are {\it sacrosanct} in 
the realm of BRST approach to gauge and/or diffeomorphism invariant theories. Second, physically, the first property (i.e. off-shell nilpotency) 
implies that these {\it fermionic} symmetry transformations are supersymmetric-type as they transform bosonic variables to fermionic 
variables and vice-versa.
Third,  the property of the absolute anticommutativity encodes the linear independence of the BRST and anti-BRST symmetry transformations.
Fourth, the absolute anticommutativity property owes its origin to the existence of the CF-type restrictions which are connected with the concepts
of gerbes [7, 8]. Fifth, as the {\it classical} gauge theory is characterized by the {\it first-class} constraints, in exactly similar fashion, the
{\it quantum} gauge and/or diffeomorphism [i.e. (anti-)BRST] 
invariant theories are characterized by the existence of the CF-type restrictions {\it within} the ambit of 
BRST formalism. Sixth, the coupled Lagrangians $L_B$ and $L_{\bar B}$ are {\it equivalent} because both of them respect BRST and 
anti-BRST symmetry transformations as is clear from Eqs. (37)-(40) provided the {\it whole} theory is considered on the submanifold of the total Hilbert 
space of the quantum variables where the CF-type restriction: $B + \bar B + i \, (\bar C \, \dot C - \dot {\bar C} \, C) = 0 $ is satisfied.\\

\section{Quantum Off-Shell Nilpotent (Anti-)BRST Symmetries of the Other Variables: ACSA}

In this section, we derive the nilpotent (anti-)BRST symmetry transformations $[s_{(a)b}]$ for {\it all} the {\it other} variables
[cf. Eqs. (32), (33)] {\it besides} the phase space variables $(x, p_x, t, p_t)$ whose (anti-)BRST symmetries have already been derived in Sec. 3
by exploiting the theoretical potential of MBTSA. To achieve the above goal, we exploit the ideas behind ACSA to BRST
formalism [25-29]. In this context, first of all, we focus on the derivation of the BRST symmetry transformations: $s_b\, B = 0, \,
s_b\,\bar B = \dot{\bar B}\,C - \bar B \, \dot C, \, s_b\,C = C\,\dot C, \, s_b\,E = \dot E\,C + E\,\dot C$
[cf. Eq. (33)]. For this purpose, we generalize the {\it ordinary} variables $[B(\tau),\, \bar B(\tau),\, C(\tau), \,E(\tau)]$ onto
a $(1, 1)$-dimensional {\it anti-chiral} super sub-manifold as follows
\begin{eqnarray}
B(\tau) \quad &\longrightarrow&  \quad {\cal B}(\tau, \bar\theta) = B(\tau) + \bar\theta\,f_1(\tau), \nonumber\\
\bar B(\tau) \quad &\longrightarrow&  \quad {\cal \bar B}(\tau, \bar\theta) = \bar B(\tau) + \bar\theta\,f_2(\tau), \nonumber\\
C(\tau) \quad &\longrightarrow&  \quad F(\tau, \bar\theta) = C(\tau) + \bar\theta\,b_1(\tau), \nonumber\\
E(\tau) \quad &\longrightarrow&  \quad \Sigma(\tau, \bar\theta) = E(\tau) + \bar\theta\,f_3(\tau), 
\end{eqnarray}
where we note that $(f_1, f_2, f_3)$ are the {\it fermionic} secondary variables and $b_1(\tau)$ is the {\it bosonic}
secondary variable because of the fermionic $({\bar\theta}^2 = 0)$ nature of the Grassmannian variable $\bar\theta$
which characterizes the {\it anti-chiral} super sub-manifold (along with the {\it bosonic} evolution parameter $\tau$).
It is elementary to note that the observation $s_b\,B = 0$ implies the following super expansion (in view of the fact that
 $\partial_{\bar \theta} \,\leftrightarrow \, s_b $), namely;
\begin{eqnarray}
{\cal B}^{(b)}\,(\tau, \bar\theta) = B(\tau) + \bar\theta\,(0) \equiv B(\tau) + \bar\theta\,(s_b \, B),
\end{eqnarray}
where the superscript $(b)$ on the {\it anti-chiral} supervariable ${\cal B}\,(\tau, \bar\theta)$ denotes that the coefficient 
of $\bar \theta$ yields the BRST symmetry transformation: $ s_b \, B = 0$ due to the trivial equality: ${\cal B}\,(\tau, \bar\theta) = B\,(\tau)$
which emerges from the observation that the Nakanishi-Lautrup auxiliary variable $B(\tau)$ is a BRST invariant quantity [cf. Eq. (33)]. In other
words, we have found out that the secondary variable $f_1 \, (\tau) = 0$ in the super expansions (41).

At this stage, we find {\it other} non-trivial BRST invariant quantities for the derivation of the secondary variables: $b_1, \, f_2,\,f_3$ of Eq. (41).
We observe that\footnote{We have specifically 
taken here $s_b\,(C\, \dot x) = 0$ for our purpose. However, one can take the general expression: $s_b\,S = C \, \dot S \,(S = x,\,p_x,\,t
,\,p_t)$ for the derivation of $b_1 (\tau) = C \, \dot C$.}: $ s_b \, ( C \, \dot {x}) = 0, \, s_b \,[\dot{\bar B}\, C  - \bar B \, \dot C] = 0, \, s_b\, [E \, \dot C + \dot E \, C] = 0$.
The basic tenets of the ACSA to BRST formalism requires that the quantities in the square brackets have to be {\it independent} of the Grassmannian variables
$\bar \theta$ when they are generalized onto a (1, 1)-dimensional super sub-manifold, namely;
\begin{eqnarray}
&F (\tau, \bar \theta) \, \dot{X}^{(h,\, ac)} \,(\tau, \bar \theta) = C \,(\tau )\, \dot x (\tau )\nonumber\\
&{\dot{\bar{\cal B}}}(\tau, \bar\theta) \,F (\tau, \bar \theta) - {\bar {\cal B}}(\tau, \bar\theta) \, \dot{F}  (\tau, \bar \theta) = \dot{\bar B}
(\tau ) \, C (\tau) - {\bar  B}(\tau ) \, \dot C (\tau) \nonumber\\
&{\Sigma }(\tau, \bar\theta) \,\dot F (\tau, \bar \theta) + \dot{\Sigma}(\tau, \bar\theta) \, {F}  (\tau, \bar \theta) = E \,
(\tau ) \, \dot C (\tau) + \dot E(\tau) \, C (\tau),
\end{eqnarray}
where ${X}^{(h,\,ac)}$ is the {\it anti-chiral} limit of the full expansion of ${X}^{(h)} \,(\tau, 
\theta, \bar \theta)$ obtained after the application of HC [cf. Eq. (28)], namely;
\begin{eqnarray}
{X}^{(h)} \,(\tau, \theta, \bar \theta) = x (\tau) + \theta \, (\bar C\, \dot x) + \bar
 \theta \,(C \, \dot {x}) + \theta\, \bar \theta \,[i\,B\,\dot x - \bar C\,\dot C\,\dot x - \bar C\,C\,\ddot x],
\end{eqnarray}
which has been obtained [cf. Eq. (28)] in Sec. 3 using the theoretical strength of MBTSA. In other words, from the top entry of Eq. (43), 
we have the following restrictions:
\begin{eqnarray}
&F (\tau, \bar \theta) \, \dot{X}^{(h, \, ac)} \,(\tau, \bar \theta) = C \,(\tau )\, \dot x(\tau )\nonumber\\
&\Longrightarrow \big[C\,(\tau) + \bar \theta \,b_1\,(\tau) \big]\, \big[ \dot x + \bar \theta \,(\dot C \, 
\dot x + C \, \ddot x)\big] = C \,(\tau )\, \dot x(\tau ).
\end{eqnarray}
From the above relationship we obtain $b_1 (\tau) = C\,\dot C$. Thus, we have the following
\begin{eqnarray}
F^{(b)} (\tau, \bar \theta) = C\,(\tau) + \bar \theta \,(C\,\dot C) \equiv C\,(\tau) + \bar \theta \,(s_b\,C),
\end{eqnarray}
where the superscript $(b)$ on the l.h.s. of the supervariable denotes that the coefficient of $\bar\theta$ is nothing but the BRST symmetry transformation
$s_b\,C$. We have to use the above super expansion in the second entry from the top in (43) to obtain the following:
\begin{eqnarray}
{\dot{\bar {\cal B}}}(\tau, \bar\theta)\,F^{(b)}(\tau, \bar\theta) - \bar{\cal B}(\tau, \bar\theta)\,{\dot F}^{(b)}(\tau, \bar\theta) = \dot{\bar B}(\tau)\,C(\tau)
- {\bar B}(\tau)\,{\dot C}(\tau).
\end{eqnarray}
In other words, we have the following equality
\begin{eqnarray}
[\dot{\bar B} + \bar\theta\,{\dot f}_2(\tau)]\,[C(\tau) + \bar\theta\,(C\,\dot C)] - [\bar B(\tau) + \bar\theta\,f_2(\tau)]\,[\dot C(\tau) + \bar\theta
\,(C\,\ddot C)] \nonumber\\ 
= \dot{\bar B}(\tau)\,C(\tau) - \bar B(\tau)\,\dot C(\tau),
\end{eqnarray} 
which yields the following condition on the secondary variable $f_2$, namely;
\begin{eqnarray}
{\dot f}_2\,C - f_2\,\dot C - \dot{\bar B} \,\dot C \, C + \bar B \,\ddot C \,C = 0.
\end{eqnarray}
It is straightforward to note that $f_2 = \dot{\bar B}\,C - \bar B\,\dot C$ satisfies the above condition in a precise manner. We point
out that the last entry (from the top) of Eq. (43) can be re-written, in view of our the super expansion in Eq. (46), as follows:
\begin{eqnarray}
\Sigma(\tau, \bar\theta)\,{\dot F}^{(b)}(\tau, \bar\theta) + \dot\Sigma(\tau, \bar\theta)\,F^{(b)}(\tau, \bar\theta) = E(\tau)\,\dot C(\tau)
+ \dot E(\tau)\,C(\tau). 
\end{eqnarray}
The substitutions of expansions from (41) and (46) lead to the following condition on the secondary variable $f_3(\tau)$ [present in the
expansion of $\Sigma(\tau, \bar\theta)$], namely;
\begin{eqnarray}
f_3\,\dot C + {\dot f}_3 \,C - E\,\ddot C \, C - \dot E\,\dot C\,C = 0,
\end{eqnarray}
which is satisfied by the choice $f_3 = E\,\dot C + \dot E\,C$. Hence, we have the following super expansions [with the BRST symmetry
transformations (33) as input], namely; 
\begin{eqnarray}
{\bar {\cal B}}^{(b)}(\tau, \bar\theta) &=& \bar B(\tau) + \bar\theta\,(\dot{\bar B}\,C - \bar B\,\dot C) \equiv \bar B(\tau) + \bar\theta\,(s_b\,\bar B),
\nonumber\\
\Sigma^{(b)}(\tau, \bar\theta) &=& E(\tau) + \bar\theta\,(E\,\dot C + \dot E\,C) \equiv E(\tau) + \bar\theta\,(s_b\,E),
\end{eqnarray}
where the coefficients of $\bar\theta$ (in view of $\partial_{\bar\theta} \leftrightarrow s_b$) are the BRST symmetry transformations (33). 
For the convenience of the readers, we have performed the
explicit computations of $f_3 = E\,\dot C + \dot E\,C$ and $f_2 = \dot {\bar B} \, C - \bar B \, \dot C$ 
in our Appendix B. It is clear that we have already computed the BRST 
transformations $s_b\,B = 0,\, s_b\,C = C\,\dot C, \, s_b\,\bar B = \dot{\bar B}\,C - \bar B \, \dot C, \, s_b\,E = E\,\dot C + \dot E\,C$
by exploiting the virtues of ACSA in Eqs. (42), (46) and (52).

We concentrate now on the derivation of the anti-BRST symmetry transformations (32) by exploiting the theoretical strength of ACSA to
BRST formalism. It is obvious that, in Sec. 3, we have already computed $s_{ab}\,S = \bar C\,\dot S \;(S = x, p_x, t, p_t)$ and $s_{ab}\,C = i\,\bar B$
by exploiting MBTSA to BRST formalism. Our objective in the present part of our section is to derive: $s_{ab}\,\bar B = 0, \, s_{ab}\,\bar C = \bar C\,\dot{\bar C},\,
s_{ab}\,B = \dot B\,\bar C - B\,\dot{\bar C},\, s_{ab}\,E = E\,\dot{\bar C} + \dot E\,\bar C$ by exploiting ACSA to BRST formalism. In this 
context, first of all, we generalize the {\it ordinary} variables onto a $(1, 1)$-dimensional {\it chiral} super sub-manifold as
\begin{eqnarray} 
\bar B(\tau) \qquad &\longrightarrow& \qquad \bar{\cal B}(\tau, \theta) = \bar B(\tau) + \theta\,{\bar f}_1(\tau), \nonumber\\
B(\tau) \qquad &\longrightarrow& \qquad {\cal B}(\tau, \theta) = B(\tau) + \theta\,{\bar f}_2(\tau), \nonumber\\
\bar C(\tau) \qquad &\longrightarrow& \qquad {\bar F}(\tau, \theta) = \bar C(\tau) + \theta\,{\bar b}_1(\tau), \nonumber\\
E(\tau) \qquad &\longrightarrow& \qquad \Sigma(\tau, \theta) = E(\tau) + \theta\,{\bar f}_3(\tau), 
\end{eqnarray}
where $({\bar f}_1,\,{\bar f}_2, {\bar f}_3)$ are the {\it fermionic} secondary variables, ${\bar b}_1(\tau)$ is a
{\it bosonic} secondary variable and the above $(1, 1)$-dimensional {\it chiral} super sub-manifold is parameterized by $(\tau, \theta)$.
It is straightforward to note that $s_{ab}\,\bar B = 0$ implies that: ${\bar {\cal B}}(\tau, \theta) = B(\tau)$ and, as a consequence,
we have ${\bar f}_1(\tau) = 0$ which leads to
\begin{eqnarray}
 {\bar{\cal B}}^{(ab)}(\tau, \theta) = \bar B(\tau) + \theta\,(0) \equiv \bar B(\tau) + \theta\,(s_{ab}\,\bar B),
\end{eqnarray}
where the superscript $(ab)$ on the {\it chiral} supervariable denotes that
we have obtained $s_{ab}\,\bar B = 0$ as the coefficient of $\theta$. The other useful and interesting anti-BRST invariant quantities of
our interest [cf. Eq. (32)] are:
\begin{eqnarray}
s_{ab}\,[\dot B\,\bar C - B\,\dot{\bar C}] = 0,   \qquad  s_{ab}\,[E\,\dot{\bar C} + \dot E \,\bar C] = 0, \qquad s_{ab}\,[\bar C\,\dot x] = 0. 
\end{eqnarray}
The quantities in the square brackets can be generalized onto the $(1, 1)$-dimensional {\it chiral} super sub-manifold. Following the
fundamental requirement(s) of ACSA to BRST formalism, {\it these} quantities must be independent of the Grassmannian variable $\theta$.
In other words, we have the following restrictions on the {\it chiral} supervariables
\begin{eqnarray}
\dot{\cal B}(\tau, \theta)\,\bar F(\tau, \theta) - {\cal B}(\tau, \theta)\,\dot{\bar F}(\tau, \theta) &=& \dot B(\tau)\,\bar C(\tau) - B(\tau)\,
\dot{\bar C}(\tau), \nonumber \\
\Sigma(\tau, \theta)\,\dot{\bar F}(\tau, \theta) + \dot\Sigma(\tau, \theta)\,\bar F(\tau, \theta) &=& E(\tau)\,\dot{\bar C}(\tau) + \dot E(\tau)\,\bar C(\tau),
\nonumber\\
\bar F(\tau, \theta)\,{\dot X}^{(h, c)}(\tau, \theta) &=& \bar C(\tau)\,\dot x(\tau),
\end{eqnarray}
where $X^{(h, c)}(\tau, \theta)$ is the {\it chiral} limit of the super expansion in Eq. (44). In other words, we have the following explicit 
expression for the supervariable $X^{(h, c)}(\tau, \theta)$, namely;
\begin{eqnarray}
X^{(h, c)}(\tau, \theta) = x(\tau) + \theta\,(\bar C\,\dot x).
\end{eqnarray}
Taking the expansions from (53) and (57), we find that the {\it last} entry of Eq. (56) yields: ${\bar b}_1(\tau) = \bar C\,\dot{\bar C}$.
Hence, we have obtained the following super expansion
\begin{eqnarray}
{\bar F}^{(ab)}(\tau, \theta) = \bar C(\tau) + \theta\,(\bar C\,\dot{\bar C}) \equiv \bar C(\tau) + \theta\,(s_{ab}\,\bar C),
\end{eqnarray} 
where the superscript $(ab)$ on the {\it chiral} supervariable on the l.h.s. denotes that it has been derived after the application
of the anti-BRST invariant restriction in (56). The coefficient $\theta$ is nothing but the anti-BRST symmetry transformation: 
$s_{ab}\,\bar C = \bar C\,\dot{\bar C}$. This equation also shows that $\partial_\theta \leftrightarrow s_{ab}$ and it leads to the
anti-BRST symmetry for $\bar C$.

We utilize now the {\it two} top entries of (56) where we use the explicit expansion for ${\bar F}^{(ab)}(\tau, \theta)$ of (58) in the following
restrictions on the supervariables, namely;
\begin{eqnarray}
\dot{\cal B}(\tau, \theta)\,{\bar F}^{(ab)}(\tau, \theta) - {\cal B}(\tau, \theta)\,\dot{\bar F}^{(ab)}(\tau, \theta) &=& {\dot B}(\tau)\,\bar C(\tau)  
- B(\tau)\,\dot{\bar C}(\tau), \nonumber \\
\Sigma(\tau, \theta)\,\dot{\bar F}^{(ab)}(\tau, \theta) + \dot\Sigma(\tau, \theta)\,{\bar F}^{(ab)}(\tau, \theta) &=& E(\tau)\,\dot{\bar C}(\tau) 
+ \dot E(\tau)\,\bar C(\tau).
\end{eqnarray} 
The substitutions from (53) and (58) lead to:
\begin{eqnarray}
\dot{\bar f}_2\,\bar C - \bar f_2\,\dot{\bar C} - \dot B\,\dot{\bar C}\,\bar C + B\,\ddot{\bar C}\,\bar C &=& 0, \nonumber \\
{\bar f}_3\,\dot{\bar C} + \dot{\bar f}_3\,{\bar C} - E\,\ddot{\bar C}\,\bar C - E\,\dot{\bar C}\,\bar C &=& 0.
\end{eqnarray}
It is straightforward, following the theoretical tricks of Appendix B, to find out the solutions for the secondary variables ${\bar f}_2(\tau)$
and ${\bar f}_3(\tau)$ which are as follows: 
\begin{eqnarray}
{\bar f}_2(\tau) = \dot B\,\bar C - B\,\dot{\bar C}, \qquad\qquad  {\bar f}_3 = E\,\dot{\bar C} + \dot E\,\bar C.
\end{eqnarray}
Substitutions of these secondary variables into the super expansions (53) leads to the determination of the anti-BRST symmetry
transformations for the variables $B(\tau)$ and $E(\tau)$ as the
coefficients of $\theta $ in the following
\begin{eqnarray}
\Sigma^{(ab)} \,(\tau, \theta) = E(\tau) + \theta \, (E\, \dot{\bar C} + \dot E \, \bar C) \,\equiv \, E(\tau) + \theta \, [s_{ab}\, E (\tau)],\nonumber\\
{\cal B}^{(ab)} \, (\tau, \theta) = B (\tau) + \theta \, (\dot B \, \bar C - B \dot {\bar C}) \, \equiv \, B (\tau) + \theta \, [s_{ab}\, B (\tau)],
\end{eqnarray}
where the superscript $(ab)$ on the {\it chiral} supervariable denotes that these supervariables have been obtained after the applications of the anti-BRST 
invariant restrictions (56). Moreover, the above observation establishes that: $s_{ab}\Leftrightarrow \partial_{\theta}$ which implies that the
nilpotency $(s_{ab}^2 = 0, \,\partial_{\theta}^2 = 0)$ properties of $s_{ab}$ and $\partial_\theta$ are connected with each-other. Thus, we have 
obtained all the anti-BRST symmetry transformations (besides the phase variables) in our Eqs. (54), (58) and (62). This completes our discussion
on the derivation of the off-shell nilpotent  and absolutely anticommuting (anti-)BRST symmetry transformations (32) and (33) within the ambit of 
ACSA to BRST formalism.\\

\vskip 0.5cm

\section{Symmetry Invariance of the Lagrangians: ACSA}

In this section, we establish the {\it equivalence} of the coupled Lagrangian $L_B$ and $L_{\bar B}$ as far as the (anti-)BRST
symmetry invariance (within the purview of ACSA to BRST formalism) is concerned. We accomplish this objective by generalizing
the {\it ordinary} Lagrangians to their counterpart {\it super} Lagrangians as 
\begin{eqnarray}  
L_{\bar B} \rightarrow {\tilde L}_{\bar B}^{(c)}(\tau, \theta) &=& {\tilde L}_f^{(c)}(\tau, \theta) - {\bar {\cal B}}^{(ab)}(\tau, \theta)\Big[{\Sigma}^{(ab)}
(\tau, \theta)\,
{\dot {\Sigma}}^{(ab)}(\tau, \theta) 
- i\,\big\{2\,{{\bar F}}^{(ab)}(\tau, \theta)\,{\dot F}^{(ab)}(\tau, \theta) \nonumber\\ 
&+& {\dot{\bar F}}^{(ab)}(\tau, \theta)\,{F}^{(ab)}(\tau, \theta)\big \}\Big] + \frac{1}{2}\,{\bar {\cal B}}^{(ab)}(\tau, \theta)\,{\bar {\cal B}}^{(ab)}
(\tau, \theta)
\nonumber\\
&-& i\,{\Sigma}^{(ab)}(\tau, \theta)\,{\Sigma}^{(ab)}(\tau, \theta)\,{\dot{\bar F}}^{(ab)}(\tau, \theta)\,{\dot F}^{(ab)}(\tau, \theta) \nonumber\\
&-& i\,{\Sigma}^{(ab)}(\tau, \theta)\,{\dot {\Sigma}}^{(ab)}(\tau, \theta)\,{{\bar F}}^{(ab)}(\tau, \theta)\,{\dot F}^{(ab)}(\tau, \theta) \nonumber\\
&-& {\dot{\bar F}}^{(ab)}(\tau, \theta)\,{\bar F}^{(ab)}(\tau, \theta)\,{\dot F}^{(ab)}(\tau, \theta)\,F^{(ab)}(\tau, \theta),
\end{eqnarray}
\begin{eqnarray}  
L_B \rightarrow {\tilde L}_B^{(ac)}(\tau, \bar\theta) &=& {\tilde L}_f^{(ac)}(\tau, \bar\theta) + {\cal B}^{(b)}(\tau, \bar\theta)
\Big[{\Sigma}^{(b)}(\tau, \bar\theta)\,{\dot {\Sigma}}^{(b)}(\tau, \bar\theta) 
- i\,\big\{2\,{\dot{\bar F}}^{(b)}(\tau, \bar\theta)\,F^{(b)}(\tau, \bar\theta) \nonumber\\ 
&+& {\bar F}^{(b)}(\tau, \bar\theta)\,{\dot F}^{(b)}(\tau, \bar\theta)\big\}\Big] + \frac{1}{2}\,{\cal B}^{(b)}(\tau, \bar\theta)\,
{\cal B}^{(b)}(\tau, \bar\theta)   \nonumber\\
&-& i\,{\Sigma}^{(b)}(\tau, \bar\theta)\,{\Sigma}^{(b)}(\tau, \bar\theta)\,{\dot{\bar F}}^{(b)}(\tau, \bar\theta)\,{\dot F}^{(b)}(\tau, \bar\theta) \nonumber\\
&-& i\,{\Sigma}^{(b)}(\tau, \bar\theta)\,{\dot {\Sigma}}^{(b)}(\tau, \bar\theta)\,{\dot{\bar F}}^{(b)}(\tau, \bar\theta)\,{F}^{(b)}(\tau, \bar\theta) \nonumber\\
&-& {\dot{\bar F}}^{(b)}(\tau, \bar\theta)\,{\bar F}^{(b)}(\tau, \bar\theta)\,{\dot F}^{(b)}(\tau, \bar\theta)\,F^{(b)}(\tau, \bar\theta),
\end{eqnarray}
where ${\tilde L}_f^{(c)}$ and ${\tilde L}_f^{(ac)}$ are the generalizations of the first-order Lagrangian $(L_f)$ to its counterpart chiral 
and anti-chiral {\it super} Lagrangians as
\begin{eqnarray}
{\tilde L}_f^{(c)}(\tau, \theta) &=& P_x^{(h, c)}(\tau, \theta)\,{\dot X}^{(h, c)}(\tau, \theta) + P_t^{(h, c)}(\tau, \theta)\,
{\dot T}^{(h, c)}(\tau, \theta) \nonumber\\
&-& \frac{\Sigma^{(ab)}(\tau, \theta)}{2}\,\Big[P_x^{(h, c)}(\tau, \theta)\,P_x^{(h, c)}(\tau, \theta)
+ 2\,m\,P_t^{(h, c)}(\tau, \theta)\Big], \nonumber\\
{\tilde L}_f^{(ac)}(\tau, \bar\theta) &=& P_x^{(h, ac)}(\tau, \bar\theta)\,{\dot X}^{(h, ac)}(\tau, \bar\theta) + P_t^{(h, ac)}(\tau, \bar\theta)\,
{\dot T}^{(h, ac)}(\tau, \bar\theta) \nonumber\\ 
&-& \frac{\Sigma^{(b)}(\tau, \bar\theta)}{2}\,\Big[P_x^{(h, ac)}(\tau, \bar\theta)\,P_x^{(h, ac)}(\tau, \bar\theta) + 2\,m\,P_t^{(h, ac)}(\tau, \bar\theta)\Big],
\end{eqnarray}
where the superscripts $(c)$ and $(ac)$ denote the {\it chiral} and {\it anti-chiral} generalizations
and the rest of the supervariables with superscripts $(b)$ and $(ab)$ have already been explained earlier in Sec. 5. The supervariables 
with superscripts $(h, c)$ and $(h, ac)$ are the {\it chiral} and {\it anti-chiral} limits of the super phase
variables $(X^{(h)}, P_x^{(h)}, T^{(h)}, P_t^{(h)})$ that have been obtained after the application of HC. Thus, these are the counterparts of the ordinary 
phase variables $(x, \,p_x,\,t,\,p_t)$ and they have been explained in 
Sec. 3. In the above equation (65), the {\it super} phase variables with superscript $(h, c)$ and $(h, ac)$ can be expressed in terms
of the {\it generic} supervariable as follows
\begin{eqnarray}
S(\tau) \quad &\rightarrow& \quad {\cal S}^{(h, c)}(\tau, \theta) = S(\tau) + \theta\,[\bar C\,\dot S (\tau)], \nonumber\\
S(\tau) \quad &\rightarrow& \quad {\cal S}^{(h, ac)}(\tau, \bar\theta) = S(\tau) + \bar\theta\,[C\,\dot S (\tau)],
\end{eqnarray}  
where the (anti-)chiral supervariables on the l.h.s. stand for the {\it super} phase variables $(X, P_x, T, P_t)$ with the
proper {\it chiral} and {\it anti-chiral} superspace coordinates $(\tau, \theta)$ and $(\tau, \bar\theta)$ as their arguments. The set of
supervariables $(X, P_x, T, P_t)$ are the generalizations of the {\it ordinary} phase variables $(x, \,p_x,\,t,\,p_t)$ to their
{\it (anti-)chiral} counterparts onto the $(1, 1)$-dimensional (anti-)chiral super submanifolds of the {\it general} $(1, 2)$-dimensional
supermanifold. It is straightforward to check that the following is true, namely;
\begin{eqnarray}
\frac{\partial}{\partial\,\theta}\,{\tilde L}_f^{(c)}(\tau, \theta) = \frac{d}{d\,\tau}[\bar C\,L_f] \quad &\Longleftrightarrow&  \quad s_{ab}\,L_f = 
\frac{d}{d\,\tau}\,[\bar C\,L_f], \nonumber\\
\frac{\partial}{\partial\,\bar\theta}\,{\tilde L}_f^{(ac)}(\tau, \bar\theta) = \frac{d}{d\,\tau}[C\,L_f] \quad &\Longleftrightarrow&  \quad s_{b}\,L_f = 
\frac{d}{d\,\tau}\,[C\,L_f].
\end{eqnarray}
In other words, we have captured the (anti-)BRST invariance of the first-order Lagrangian $(L_f)$ in view of the mappings:
$s_b \leftrightarrow \partial_{\bar\theta}, \, s_{ab} \leftrightarrow \partial_{\theta}$. Since in the {\it ordinary} space, the 
(anti-) BRST symmetry transformations acting on $L_f$ produce the {\it total} derivatives [cf. Eq. (67)], the action 
integral $S = \int^{+\infty}_{-\infty}\,d\,\tau\,L_f$ remains invariant under the transformations $s_{(a)b}$.

At this stage, we focus on the (anti-)BRST invariance of the coupled Lagrangians $L_B$ and $L_{\bar B}$ [cf. Eqs. (38), (37)]. 
We can express {\it these} invariances within the ambit of ACSA (in view of the mappings: $s_b \leftrightarrow \partial_{\bar\theta}, \,
s_{ab} \leftrightarrow \partial_\theta$), namely;
\begin{eqnarray}
\frac{\partial}{\partial\,\theta}\,{\tilde L}_{\bar B}^{(c)}(\tau, \theta) = \frac{d}{d\,\tau}\,\Big[\bar C\,L_f - e\,\dot e\,\bar B\,\bar C
- e^2\,\bar B\,\dot{\bar C} + {\bar B}^2\,\bar C  - i\,\bar B\,\dot{\bar C}\,\bar C\,C  \Big] = s_{ab}\,L_{\bar B},
\end{eqnarray}
\begin{eqnarray} 
\frac{\partial}{\partial\,\bar\theta}\,{\tilde L}_B^{(ac)}(\tau, \bar\theta) = \frac{d}{d\,\tau}\,\Big[C\,L_f + e\,\dot e\,B\,C 
+ e^2\,B\,\dot C + B^2\,C - i\,B\,\bar C\,\dot C\,C \Big] = s_b\,L_B,
\end{eqnarray}
where the {\it super} Lagrangians ${\tilde L}_{\bar B}^{(c)}(\tau, \theta)$ and ${\tilde L}_B^{(ac)}(\tau, \bar\theta)$ have been 
already quoted in Eqs. (63) and (64). It is interesting to note that the r.h.s. of (68) and (69) are {\it same} as we have found
in the {\it ordinary} space [cf. Eq. (38), (37)]. To prove the {\it equivalence} of the Lagrangians $L_B$ and $L_{\bar B}$ w.r.t.
the (anti-)BRST symmetry transformations [cf. Eqs. (32), (33)] within the purview of ACSA, we generalize the {\it ordinary} Lagrangians
$L_B$ and $L_{\bar B}$ as follows
\begin{eqnarray}  
L_B \rightarrow {\tilde L}_B^{(c)}(\tau, \theta) &=& {\tilde L}_f^{(c)}(\tau, \theta) + {\cal B}^{(ab)}(\tau, \theta)
\Big[\Sigma^{(ab)}(\tau, \theta)\,{\dot \Sigma}^{(ab)}(\tau, \theta) 
- i\,\{2\,{\dot{\bar F}}^{(ab)}(\tau, \theta)\,F^{(ab)}(\tau, \theta) \nonumber\\ 
&+& {\bar F}^{(ab)}(\tau, \theta)\,{\dot F}^{(ab)}(\tau, \theta)\}\Big] + \frac{1}{2}\,{\cal B}^{(ab)}(\tau, \theta)\,
{\cal B}^{(ab)}(\tau, \theta)
\nonumber \\
&-& i\,\Sigma^{(ab)}(\tau, \theta)\,\Sigma^{(ab)}(\tau, \theta)\,{\dot{\bar F}}^{(ab)}(\tau, \theta)\,{\dot F}^{(ab)}(\tau, \theta) \nonumber\\
&-& i\,\Sigma^{(ab)}(\tau, \theta)\,{\dot \Sigma}^{(ab)}(\tau, \theta)\,{\dot{\bar F}}^{(ab)}(\tau, \theta)\,{F}^{(ab)}(\tau, \theta) \nonumber\\
&-& {\dot{\bar F}}^{(ab)}(\tau, \theta)\,{\bar F}^{(ab)}(\tau, \theta)\,{\dot F}^{(ab)}(\tau, \theta)\,F^{(ab)}(\tau, \theta),
\end{eqnarray}   
 \begin{eqnarray}  
L_{\bar B} \rightarrow {\tilde L}_{\bar B}^{(ac)}(\tau, \bar\theta) &=& {\tilde L}_f^{(ac)}(\tau, \bar\theta) - {\bar{\cal B}}^{(b)}(\tau,
 \bar\theta)\Big[\Sigma^{(b)}(\tau, \bar\theta)\,{\dot \Sigma}^{(b)}(\tau, \bar\theta) - i\,\big\{2\,{{\bar F}}^{(b)}(\tau, \bar\theta)\,
{\dot F}^{(b)}(\tau, \bar\theta)
 \nonumber\\ 
&+& {\dot{\bar F}}^{(b)}(\tau, \bar\theta)\,{F}^{(b)}(\tau, \bar\theta)\big \}\Big] + \frac{1}{2}\,{\bar {\cal B}}^{(b)}(\tau, \bar\theta)\,
{\bar {\cal B}}^{(b)}(\tau, \bar\theta)
\nonumber\\
&-& i\,\Sigma^{(b)}(\tau, \bar\theta)\,\Sigma^{(b)}(\tau, \bar\theta)\,{\dot{\bar F}}^{(b)}(\tau, \bar\theta)\,{\dot F}^{(b)}(\tau, \bar\theta)
 \nonumber\\
&-& i\,\Sigma^{(b)}(\tau, \bar\theta)\,{\dot \Sigma}^{(b)}(\tau, \bar\theta)\,{{\bar F}}^{(b)}(\tau, \bar\theta)\,{\dot F}^{(b)}(\tau,
 \bar\theta) \nonumber\\
&-& {\dot{\bar F}}^{(b)}(\tau, \bar\theta)\,{\bar F}^{(b)}(\tau, \bar\theta)\,{\dot F}^{(b)}(\tau, \bar\theta)\,F^{(b)}(\tau, \bar\theta),
\end{eqnarray}
where {\it all} the notations and symbols have already been explained earlier. To find out the result of the operations of $s_b$ on $L_{\bar B}$ and $s_{ab}$
on $L_B$, we observe the following (in view of the mappings: $s_b \leftrightarrow \partial_{\bar\theta},\,s_{ab} \leftrightarrow \partial_{\theta}$), namely;
\begin{eqnarray}
\frac{\partial}{\partial\,\theta}\,{\tilde L}_B^{(c)}(\tau, \theta) &=& \frac{d}{d\,\tau}\,\Big[\bar C\,L_f + E\,\dot E\,(i\,\dot{\bar C}\,
\bar C\,C + B\,\bar C) + E^2\,(i\,\dot{\bar C}\,
\bar C\,\dot C + B \,\dot{\bar C})  \nonumber\\ 
&+& {B}^2\,\bar C + i\,(2\,B - \bar B)\,\dot{\bar C}\,\bar C\,C\Big] \nonumber\\
&+& \big[B+ \bar B + i\,(\bar C\,\dot C - \dot{\bar C}\,C)\big]\,(2\,i\,\dot{\bar C}\,\bar C\,\dot C 
- E\,\dot E\,\dot{\bar C} - 2\,B\,\dot{\bar C}  + i\,\ddot{\bar C}\,\bar C\,C) \nonumber\\
&-& \frac{d}{d\,\tau}\big[B+ \bar B + i\,(\bar C\,\dot C - \dot{\bar C}\,C)\big]\,( B\,\bar C + E^2\,\dot{\bar C}) \equiv s_{ab}\,L_B, 
\end{eqnarray}
\begin{eqnarray}
\frac{\partial}{\partial\,\bar\theta}\,{\tilde L}_{\bar B}^{(ac)}(\tau, \bar\theta) &=& \frac{d}{d\,\tau}\,\Big[C\,L_f 
 - E\,\dot E\,(i\,\bar C\,\dot C\, C + \bar B\,C)- E^2\,(i\,\dot{\bar C}\,\dot C\, C + \bar B \,\dot C) \nonumber\\ 
&+& {\bar B}^2\,C + i\,(2\,\bar B - B)\,\bar C\,\dot C\, C \Big] \nonumber\\
&+& \big[B+ \bar B + i\,(\bar C\,\dot C - \dot{\bar C}\,C)\big]\,\big [i\,\bar C\,\ddot C\,C + 2\,i\,
\dot{\bar C}\,\dot C\, C - 2\,\bar B\,\dot C + E\,\dot E\,\dot C \big ] \nonumber\\
&+& \frac{d}{d\,\tau}\big[i\,(\bar C\,\dot C - \dot{\bar C}\,C) + B + \bar B \big]\,(E^2\,\dot C - \bar B\,C) \equiv s_b\,L_{\bar B}, 
\end{eqnarray}
within the framework of ACSA. It is self-evident, from the r.h.s. of (72) and (73), that we have the 
BRST invariance of $L_{\bar B}$ and anti-BRST invariance of $L_B$ if and only if our whole theory is considered on
the submanifold of the Hilbert space of quantum variables where the 
CF-type restriction: $B + \bar B + i\,(\bar C\,\dot C - \dot{\bar C}\,C) = 0$ is satisfied.

We end this section with the following crucial remarks. First of all, we have captured the BRST 
and anti-BRST invariance of $L_B$ and $L_{\bar B}$, respectively, in the terminology of ACSA on the (anti-)chiral super submanifolds
[cf. Eqs. (68), (69)]. Second, we have {\it also} demonstrated the anti-BRST invariance of $L_B$ {\it and} BRST invariance of $L_{\bar B}$
in the superspace formalism [cf. Eqs. (72), (73)] where the theoretical techniques of ACSA have played very important roles. Third,
we have also expressed the (anti-)BRST invariance of the first-order Lagrangian $L_f$ in Eq. (67).
Finally, we have proven the {\it equivalence } of $L_B$ and $L_{\bar B}$ within the framework of ACSA in the Eqs. (68), (69), (72) and (73).\\

\section{Nilpotency and Absolute Anticommutativity Properties of the (Anti-)BRST Charges: ACSA}

Our present section is divided into two subsections. In subsection 7.1, we discuss the off-shell nilpotency 
and absolute anticommutativity of the conserved (anti-)BRST charges in the {\it ordinary} space. Our subsection 7.2
deals with the {\it above} properties within the realm of ACSA to BRST formalism. In other words, we capture the off-shell
nilpotency and absolute anticommutativity of the conserved fermionic (anti-)BRST charges in the superspace by taking the theoretical inputs from ACSA.

\subsection{Nilpotency and Anticommutativity: Ordinary Space}

The perfect symmetry invariance of $L_{\bar B}$ under the anti-BRST symmetry transformations 
[cf. Eq. (38)] {\it and} $L_B$ under the BRST symmetry transformations [cf. Eq. (37)] allow us to compute the Noether
conserved charges by using the standard techniques of Noether's theorem (applied to the 
action integrals corresponding to the Lagrangians $L_{\bar B}$ and $L_B$) as 
\begin{eqnarray}
Q_{\bar B} &=& \frac{\bar C\,E}{2}\,(p_x^2 + 2\,m\,p_t) + {\bar B}^2\,\bar C  - i\,\bar B\,\dot{\bar C}\,\bar C\,C 
- \bar B\,E\,\dot E\,\bar C - \bar B\,E^2\,\dot{\bar C} , \nonumber\\
Q_B &=& \frac{C\,E}{2}\,(p_x^2 + 2\,m\,p_t) + B\,E^2\,\dot C  - i\,B\,\bar C\,\dot C\,C + B\,E\,\dot E\,C + B^2\,C ,
\end{eqnarray}
where conserved $({\dot Q}_{\bar B} = 0, \, {\dot Q}_B = 0)$ (anti-)BRST charges are denoted by $Q_{(\bar B)B}$.
The conservation law $({\dot Q}_{\bar B} = 0, \, {\dot Q}_B = 0)$ can be proven by using the EL-EOMs derived from the coupled Lagrangians $L_{\bar B}$
and $L_B$. For readers' convenience, we prove the conservation $({\dot Q}_B = 0)$ of the BRST charge by using the 
EL-EOMs derived from $L_B$ in our Appendix C.

First of all, we concentrate on the proof of the off-shell nilpotency properties of the (anti-)BRST charges $Q_{(\bar B)B}$.
In this context, we note that the following EL-EOMs w.r.t. the variable $E$ from $L_{\bar B}$ and $L_B$, respectively, yield the following:
\begin{eqnarray}
\dot{\bar B}\,E - i\,E\,\dot{\bar C}\,\dot C + i\,E\,\bar C\,\ddot C - \frac{1}{2}\,(p_x^2 + 2\,m\,p_t) &=& 0, \nonumber\\
\dot B\,E + i\,E\,\dot{\bar C}\,\dot C - i\,E\,\ddot{\bar C}\,C + \frac{1}{2}\,(p_x^2 + 2\,m\,p_t) &=& 0.
\end{eqnarray}
The above equations can be used to recast $Q_{\bar B}$ and $Q_B$ as follows:
\begin{eqnarray} 
Q_{\bar B} ^{(1)} &=& E^2\,(\dot{\bar{B}}\,{\bar{C}} - {\bar{B}}\,\dot{\bar{C}} + i\,\dot {\bar C}\,\bar{C}\,\dot C) 
- i\,\bar{B}\,\dot{\bar C}\,{\bar{C}}\,C - \bar{B}\,E\,\dot E\, \bar{C} + \bar{B}^2\, \bar{C}, \nonumber\\
Q_B ^{(1)} &=& E^2\,(B\,\dot C - \dot B\,C  - i\,\dot {\bar C}\,\dot C\,C) -  i\,B\,\bar C\,\dot C\,C + B\,E\,\dot E\, C + B^2\, C,
\end{eqnarray}
Using the following EL-EOMs w.r.t. the variables $C$ and $\bar B$, respectively, from $L_{\bar B}$, namely;
\begin{eqnarray}
 &&i\,\bar B\,\dot{\bar C} + 2\,i\,\dot{\bar B}\,\bar C - 3\,i\,E\,\dot E\,\dot{\bar C} - i\,E^2\,\ddot{\bar C} - i\,{\dot E}^2\,\bar C 
- i\,E\,\ddot E\,\bar C + \ddot{\bar C}\,\bar C\,C + 2\,\dot{\bar C}\,\bar C\,\dot C = 0, \nonumber\\
&&\bar B = E\,\dot E - i\,(2\,\bar C\,\dot C + \dot{\bar C}\,C),
\end{eqnarray}
we obtain the following {\it exact} and interesting expression for the anti-BRST charge:
\begin{eqnarray}
Q_{\bar B}^{(1)} \quad \longrightarrow  \quad Q_{\bar B}^{(2)} &=& E^2\,(\dot{\bar{B}}\,{\bar{C}} - {\bar{B}}\,\dot{\bar{C}} 
+ i\,\dot {\bar C}\,\bar{C}\,\dot C) + i\,E^2\,\ddot{\bar C}\,\bar C\,C + 2\,i\,E\,\dot E\,\dot{\bar C}\,\bar C\,C \nonumber\\
&\equiv& s_{ab}\,[i\,E^2\,(\bar C\,\dot C - \dot{\bar C}\,C)].
\end{eqnarray} 
At this juncture, we apply the basic principle behind the relationship between the continuous symmetry transformations (e.g. $s_{ab}$)
and its generator $[Q_{\bar B}^{(2)}]$ which implies that:
\begin{eqnarray}
s_{ab}\,Q_{\bar B}^{(2)} = -\,i\,\{Q_{\bar B}^{(2)}, Q_{\bar B}^{(2)}\} = 0 \quad \Rightarrow \quad [Q_{\bar B}^{(2)}]^2 = 0 \quad \Leftrightarrow 
\quad s_{ab}^2 = 0.
\end{eqnarray}
Thus, we observe that the off-shell nilpotency $([Q_{\bar B}^{(2)}]^2 = 0)$ of the anti-BRST charge $Q_{\bar B}^{(2)}$ and the anti-BRST symmetry 
transformations $(s_{ab})$
are {\it inter-related}. Thus, we have proven the off-shell nilpotency of the anti-BRST charge $Q_{\bar B}^{(2)}$. In exactly 
similar fashion, we exploit the following EL-EOMs w.r.t. the variables $\bar C$ and $B$ from the Lagrangian $L_B$
\begin{eqnarray}
&&i\,B\,\dot C + 2\,i\,\dot B\,C + 3\,i\,E\,\dot E\,\dot C + i\,E^2\,\ddot C + i\,{\dot E}^2\,C + i\,E\,\ddot E\,C + \bar C\,\ddot C\,C
+ 2\,\dot{\bar C}\,\dot C\,C = 0, \nonumber\\
&& B = - E\,\dot E + i\,(2\,\dot{\bar C}\,C + \bar C\,\dot C),
\end{eqnarray}
to recast the BRST charge $Q_B^{(1)}$ into another interesting form (i.e. $Q_B^{(2)}$) as
\begin{eqnarray}
Q_B^{(1)} \quad \rightarrow \quad Q_B^{(2)} &=& E^2\,(B\,\dot C - i\,\dot{\bar C}\,\dot C\,C - \dot B\,C) - 2\,i\,E\,\dot E\,\bar C\,\dot C\,C 
- i\,E^2\,\bar C\,\ddot C\,C
 \nonumber\\
&\equiv& s_b\,[i\,E^2\,(\dot{\bar C}\,C - \bar C\,\dot C)] 
\end{eqnarray}
which turns out to be an {\it exact} quantity w.r.t. $s_b$. Thus, we find that we have the following:
\begin{eqnarray}
s_{b}\,Q_{B}^{(2)} = -\,i\,\{Q_{B}^{(2)}, Q_{B}^{(2)}\} = 0 \quad \Rightarrow \quad [Q_{B}^{(2)}]^2 = 0 \quad \Leftrightarrow 
\quad s_{b}^2 = 0.
\end{eqnarray}
In other words, we have proven the off-shell nilpotency $([Q_B^{(2)}]^2 = 0)$ of the BRST charge $Q_B^{(2)}$. Once again, we find that off-shell
nilpotency $(s_{b}^2 = 0)$ of the BRST symmetry transformations and the off-shell nilpotency $([Q_{B}^{(2)}]^2 = 0)$ are intertwined an intimate manner.

We now focus on the proof of the absolute anticommutativity property of the BRST charge with the anti-BRST charge  and vice-versa.
First of all, let us focus on the BRST charge $Q_{B}^{(2)}$ [cf. Eq. (81)]. Using the CF-type restriction: $B + \bar B + i\,
(\bar C\,\dot C - \dot {\bar C}\, C) = 0$, we can easily check the following transformation:
\begin{eqnarray}
Q_{B}^{(2)} \quad \longrightarrow \quad Q_{B}^{(3)} &=& E^2 \,(\dot{\bar B}\,C - 2\,i\,\dot{\bar C}\,\dot C\,C - \bar B\,\dot C ) 
- 2\,i\,E\,\dot E\,\bar C\,\dot C\,C
\nonumber\\
&\equiv& s_{ab}\,[i\,E^2\,\dot C\,C].
\end{eqnarray}
In other words we have been able to express the above BRST charge as an {\it exact} form w.r.t. the anti-BRST symmetry transformations $(s_{ab})$.
This is an interesting observation because using the relationship between the continuous symmetry transformations and their generators, we can
obtain the following from (83), namely;
\begin{eqnarray}
s_{ab}\,Q_B^{(3)} = -i\,\{Q_B^{(3)}, Q_{\bar B}^{(3)}\} = 0 \quad \Leftrightarrow \quad s_{ab}^2 = 0.
\end{eqnarray}
Thus we have been able to demonstrate that the absolute anticommutativity of the BRST charge {\it with} the anti-BRST charge is connected with 
the off-shell nilpotency $(s_{ab}^{2} = 0)$ of the anti-BRST symmetry transformation $(s_{ab})$. In exactly similar fashion, we can have a different form 
of the anti-BRST charge $Q_{\bar B}^{(2)}$ [cf. Eq. (78)] by using the CF-type restriction:  $B + \bar B + i\,
(\bar C\,\dot C - \dot {\bar C}\, C) = 0$. In other words, we have the following interesting transformation: 
\begin{eqnarray}
Q_{\bar B}^{(3)} \quad \longrightarrow \quad  Q_{\bar B}^{(3)} &=& E^2\,(B\,\dot{\bar C} + 2\,i\,\dot{\bar C}\,\bar C\,\dot C - \dot B\,\bar C ) + 
2\,i\,E\,\dot E\,\dot{\bar C}\,\bar C\,C \nonumber\\
&\equiv& s_b\,[i\,E^2\, \dot{\bar C}\,\bar C]. 
\end{eqnarray}
It is straightforward to note that we have the following relationship:
\begin{eqnarray}
s_b\,Q_{\bar B}^{(3)} = -\,i\,\{Q_{\bar B}^{(3)}, Q_B^{(3)}\} = 0 \quad \Leftrightarrow \quad s_b^2 = 0.
\end{eqnarray}
In other words, we point out that the absolute anticommutativity of the anti-BRST charge {\it with} the BRST
charge is intimately connected with the off-shell nilpotency $(s_b^2 = 0)$ of the BRST symmetry transformations
$(s_b)$. This completes our discussions on the off-shell nilpotency and absolute anticommutativity of 
the conserved (anti-)BRST charges in the {\it ordinary} space. In a subtle manner, the observations 
in (83) and (85) prove the validity of the CF-type restriction:  $B + \bar B + i\,
(\bar C\,\dot C - \dot {\bar C}\, C) = 0$ {\it on} our theory.

\subsection{Nilpotency and Anticommutativity: ACSA}

The key observations of subsection $7.1$ can be translated into the {\it superspace} by using the basic
terminology of ACSA. Keeping in our mind the mappings $\partial_{\bar\theta} \leftrightarrow s_b, \, 
\partial_{\theta} \leftrightarrow s_{ab}$, we note that the (anti-)BRST charges $Q_{(\bar B)B}$ [cf. 
Eqs. (78), (81)] can be expressed as:
\begin{eqnarray}
Q_{\bar B} = \frac{\partial}{\partial\,\theta}\,\Big[i\,E^{(ab)}(\tau, \theta)\,E^{(ab)}(\tau, \theta)\,
\big\{{\bar F}^{(ab)}(\tau, \theta)\,{\dot F}^{(ab)}(\tau, \theta) - \dot{\bar F}^{(ab)}(\tau, \theta)\,F^{(ab)}
(\tau, \theta)\big\}\Big] \nonumber\\
= \int d\,\theta \,\Big[i\,E^{(ab)}(\tau, \theta)\,E^{(ab)}(\tau, \theta)\,
\big\{{\bar F}^{(ab)}(\tau, \theta)\,{\dot F}^{(ab)}(\tau, \theta) - \dot{\bar F}^{(ab)}(\tau, \theta)\,F^{(ab)}
(\tau, \theta)\big\}\Big]
\end{eqnarray}    
\begin{eqnarray}
Q_{B} = \frac{\partial}{\partial\,\bar\theta}\,\Big[i\,E^{(b)}(\tau, \bar\theta)\,E^{(b)}(\tau, \bar\theta)\,
\big\{\dot{\bar F}^{(b)}(\tau, \bar\theta)\,{F}^{(b)}(\tau, \bar\theta) - {\bar F}^{(b)}(\tau, \bar\theta)\,{\dot F}^{(b)}
(\tau, \bar\theta)\big\}\Big] \nonumber\\
= \int d\,\bar\theta \,\Big[i\,E^{(b)}(\tau, \bar\theta)\,E^{(b)}(\tau, \bar\theta)\,
\big\{\dot{\bar F}^{(b)}(\tau, \bar\theta)\,{F}^{(b)}(\tau, \bar\theta) - {\bar F}^{(b)}(\tau, \bar\theta)\,{\dot F}^{(b)}
(\tau, \bar\theta)\big\}\Big].
\end{eqnarray} 
It is straightforward to observe that we have the following:
\begin{eqnarray}
\frac{\partial}{\partial\,\theta}\,Q_{\bar B} = 0 \quad &\Leftrightarrow& \quad s_{ab}\,Q_{\bar B} = 0 \quad \Leftrightarrow 
\quad Q_{\bar B}^2 = 0 \quad \Leftrightarrow \quad \partial_\theta^2 = 0, \nonumber\\
\frac{\partial}{\partial\,\bar\theta}\,Q_{B} = 0 \quad &\Leftrightarrow& \quad s_{b}\,Q_{B} = 0 \quad \Leftrightarrow 
\quad Q_B^2 = 0 \quad \Leftrightarrow \quad \partial_{\bar\theta}^2 = 0.
\end{eqnarray}
Thus, the off-shell nilpotency of the (anti-)BRST charges is connected with the nilpotency 
$(\partial_\theta^2 = 0,\,\partial_{\bar\theta}^2 = 0)$ of the translational generators $(\partial_\theta, 
\partial_{\bar\theta})$ along the Grassmannian directions of the {\it chiral} and {\it anti-chiral} $(1, 1)$-dimensional 
super sub-manifolds. This observation is consistent with our discussion of the nilpotency property
in the {\it ordinary} space if we remember the mappings: $s_b \leftrightarrow \partial_{\bar\theta},\,s_{ab}
\leftrightarrow \partial_\theta$ [14-16].

As far as the absolute anticommutativity property is concerned, we note that the expressions of the (anti-)BRST
charges in (85) and (83) can be translated into the {\it superspace} where we can exploit the theoretical tools of
ACSA. To accomplish this goal, we keep in our knowledge the mappings: $s_b \leftrightarrow \partial_{\bar\theta},\,s_{ab}
\leftrightarrow \partial_\theta$ to recast the expressions (85) and (83) as:
\begin{eqnarray}
Q_{\bar B}^{(3)} &=& \frac{\partial}{\partial\,\bar\theta}\,\Big[i\,E^{(b)}(\tau, \bar\theta)\,E^{(b)}(\tau, \bar\theta)
\,\dot{\bar F}^{(b)}(\tau, \bar\theta)\,{\bar F}^{(b)}(\tau, \bar\theta)\Big] \nonumber\\
&\equiv& \int d\,\bar\theta \,\Big[i\,E^{(b)}(\tau, \bar\theta)\,E^{(b)}(\tau, \bar\theta)
\,\dot{\bar F}^{(b)}(\tau, \bar\theta)\,{\bar F}^{(b)}(\tau, \bar\theta)\Big],\nonumber\\
Q_B^{(3)} &=& \frac{\partial}{\partial\,\theta}\,\Big[- i\,E^{(ab)}(\tau, \theta)\,E^{(ab)}(\tau, \theta)
\,{\dot F}^{(ab)}(\tau, \theta)\,{F}^{(ab)}(\tau, \theta)\Big] \nonumber\\
&\equiv& \int d\,\theta\,\Big[- i\,E^{(ab)}(\tau, \theta)\,E^{(ab)}(\tau, \theta)
\,\dot{F}^{(ab)}(\tau, \theta)\,{F}^{(ab)}(\tau, \theta)\Big]. 
\end{eqnarray}   
It is now straightforward to check that the following are true, namely; 
\begin{eqnarray}
\partial_{\bar\theta}\,Q_{\bar B}^{(3)} = 0 \quad &\Leftrightarrow& \quad s_b\,Q_{\bar B}^{(3)} = 0 \quad \Leftrightarrow \quad
\{Q_{\bar B}^{(3)}, Q_{B}^{(3)}\} = 0 \quad \Leftrightarrow \quad \partial_{\bar\theta}^2 = 0, \nonumber\\
\partial_{\theta}\,Q_{B}^{(3)} = 0 \quad &\Leftrightarrow& \quad s_{ab}\,Q_{B}^{(3)} = 0 \quad \Leftrightarrow \quad
\{Q_{B}^{(3)}, Q_{\bar B}^{(3)}\} = 0 \quad \Leftrightarrow \quad \partial_{\theta}^2 = 0,
\end{eqnarray}
which establishes the fact that the ACSA to BRST formalism distinguishes between the {\it two} types
of absolute anticommutativity properties. In other words, we observe that the absolute anticommutativity of the BRST charge {\it with}
the anti-BRST charge is connected with the nilpotency $(\partial_{\theta}^2 = 0)$ of the translational generator $(\partial_\theta)$
along the Grassmannian direction of the (1, 1)-dimensional {\it chiral} super sub-manifold. On the contrary, the absolute 
anticommutativity of the anti-BRST
charge {\it with} the BRST charge is connected with the nilpotency $(\partial^2_{\bar\theta} = 0)$ of the 
translational generator $(\partial_{\bar \theta})$ along
the Grassmannian direction of the (1, 1)-dimensional {\it anti-chiral} super sub-manifold.

\section{Conclusions}

In our present endeavor, we have purposely taken a reparameterization invariant NR and NSUSY system so that we 
could discuss theoretical aspects  that are different from our earlier works on the NSUSY relativistic 
{\it scalar} and SUSY relativistic {\it spinning} particles [23, 24]. We have demonstrated, however, in our present 
investigation that $(i)$ the CF-type restriction, and $(ii)$ the sum of gauge-fixing and Faddeev-Popov ghost
terms are {\it same} for our present NR and NSUSY system as have been shown by us for the relativistic particles (in
our earlier works [23, 24]). The above observations are interesting results of our present investigation which 
establish the {\it universality} of the (anti-)BRST invariant CF-type restriction for the 1D diffeomorphism invariant
(i.e. reparameterization) theories.

The CF-type restriction(s) are the hallmark of a {\it quantum} theory that is BRST-{\it quantized}. In fact, for a 
D-dimensional diffeomorphism invariant theory, it has been shown [22, 33] that the {\it universal} CF-type
restrictions for a BRST-{\it quantized} theory is: $B_{\mu} + {\bar B}_\mu + i\, ({\bar C}^{\rho}\,\partial_\rho\,C_\mu
 + {C}^{\rho}\,\partial_\rho\, \bar C_\mu) = 0$ where $\mu, \rho = 0, 1, 2,...D-1$, $B_\mu$ and $\bar B_\mu$ are 
 the Nakanishi-Lautrup auxilary fields and the (anti-)ghost fields $(\bar C_\mu)\,C_\mu$ correspond to the 
D-dimensional diffeomorphism parameter $\epsilon_{\mu}(x)$ in the infinitesimal transformation: $x_\mu\rightarrow 
x_\mu' = x_\mu - \epsilon_{\mu}(x)$. The {\it universality} of the above CF-type restriction implies that, for our
1D diffeomorphism (i.e. reparameterization) invariant theory, the CF-type restriction is: $B + \bar B + i\,
(\bar C\,\dot C - \dot {\bar C}\, C) = 0$. This is what we have obtained from various theoretical tricks in  our
present endeavor. The existence of the CF-type restriction is very fundamental to a BRST-quantized theory as it is connected
with the geometrical objects called gerbes [7, 8]. Physically, the existence of the CF-type restriction leads to the independent
nature of the BRST and anti-BRST symmetries (and corresponding conserved charges) at the {\it quantum} level 
(that are connected with a given {\it classical} local symmetry).

Our present work (and earlier works [23, 24]) can be generalized to the cases of (super)string and gravitational theories which are also 
diffeomorphism invariant. In fact, in our earlier work on a bosonic string theory [34], we have shown the existence 
of the CF-type restriction in the context of its BRST quantization and it has turned out to be the 2D version of the {\it universal}
CF-type restriction for the D-dimensional diffeomorphism invariant theory. It is gratifying to pinpoint the fact that
we have derived the CF-type restrictions: $B^a + \bar B^a + i \; \big (\bar C^m \; \partial_m\; C^a +  C^m \; \partial_m\; \bar C^a \big ) = 0$
(with $a, m = 0, 1$) for a model of bosonic string theory [34].   
This has happened because the bosonic string theory has the 
2D diffeomorphism invariance on the 2D world-sheet. We have applied the beautiful blend of MBTSA and ACSA to derive {\it all}
the (anti-)BRST symmetries as well as the 2D version of the CF-type restriction in the case of a bosonic string theory of our interest [35].
In our present investigation, we have utilized only {\it two} and/or {\it one} Grassmannian variables because there are {\it only}
two nilpotent symmetries in the theory. If a theory is endowed with the nilpotent (anti-)BRST as well as (anti-)co-BRST symmetries, then, we have to
invoke {\it four} number of Grassmannian variables. We are currently exploring such kinds of possibilities.\\

\vskip 0.3cm

\begin{center}
{\bf Appendix A: The CF-Type Restriction from $L_B \equiv L_{\bar B}$ }\\
\end{center}

\vskip 0.3cm

\noindent
In this Appendix, we provide the step-by-step derivation of the CF-type restriction by requiring the equivalence of the coupled Lagrangian $L_B$ and $L_{\bar B}$
[cf. Eq. (35)]. A close look at them demonstrates that if we demand $L_B \equiv L_{\bar B}$, the terms that are common would cancel out. For instance, we have
cancellations of terms $L_f,\, -i \, E^2 \, \dot {\bar C} \, \dot C$ and $ - \, \dot {\bar C} \, \bar C \, \dot C \, C$ that are present {\it both} in $L_B$ 
and $L_{\bar B}$.
Thus, we are left with the following equality:
\[
\frac{B^2}{2} + B\,[E \, \dot E - i \, (2 \, \dot{\bar C} \, C + \bar C \, \dot C)] - i \, E \, \dot E \, \dot {\bar C}\, C
\]
\[
\equiv \frac{\bar B^2}{2} + \bar B\,[E \, \dot E - i \, (2 \, {\bar C} \, \dot C + \dot{\bar C} \, C)] - i \, E \, \dot E \, {\bar C}\, \dot C. 
\eqno (A.1)
\]
From the above equation, it is evident that we have
\[
E \, \dot E \, [ B + \bar B + i \, (\bar C \, \dot C - \dot {\bar C} \, C)],
\eqno (A.2)
\]
on the l.h.s. when we bring {\it all} the terms from 
the r.h.s. to the l.h.s.. At this stage, excluding (A.2), the {\it left-over} terms on the l.h.s. and the r.h.s. are
\[
\frac{B^2}{2} - \frac{\bar B^2}{2} - 2 \, i \, B \, \dot {\bar C}\, C - i \, B \, \bar C \, \dot C 
- 2\, i \, \bar B\, \bar C \dot C - i\, \bar B \dot {\bar C} \, C = 0.
\eqno(A.3)
\]
The above equation can be expressed as:
\[
\frac{B^2}{2} - \frac{\bar B^2}{2} - i \,(B + \bar B)\,\dot{\bar C}\, C - i\, (B + \bar B)\,\bar C \, \dot C - i\ B \, \dot {\bar C}\,C 
- i\, \bar B \, \bar C \, \dot C = 0.
\eqno(A.4)
\]
The re-arrangements of the terms produce the following
\[
\frac{B^2}{2} - \frac{\bar B^2}{2} - i\,[B + \bar B + i \, (\bar C \, \dot C - \dot {\bar C} \, C)]\,\dot {\bar C}\,C - i \,[B + \bar B + i \, (\bar C \, \dot C - \dot {\bar C} \, C)]\, \bar C\, \dot C
\]
\[- i\, B \dot {\bar C}\, C - i\, \bar B \, \bar C\, \dot C = 0.
\eqno(A.5)
\]
Taking into account $(A.2)$, we have the following
\[
\big[E\, \dot E - i \, \dot {\bar C}\, C - i \, \bar C \, \dot C \big]\,\big[B + \bar B + i \, (\bar C \, \dot C - \dot {\bar C} \, C)\big] + \frac{B^2}{2} - \frac{\bar B^2}{2} 
\]
\[- i\ B \, \dot {\bar C}\,C - i\, \bar B \, \bar C \, \dot C = 0.
\eqno(A.6)
\]
Substituting for $ -\, i \, B\, \dot {\bar C} \,C = -\,\frac{i}{2} \, B\, \dot {\bar C} \,C  - \,\frac{i}{2} \, B\, \dot {\bar C} \,C$ 
and $- \, i\, \bar B \bar C \dot C = -\, \frac{i}{2}\, \bar B \, \bar C \, \dot C - \frac{i}{2} \, \bar B \, \bar C \, \dot C$ and re-arranging 
the terms, we end up with the following {\it final} result:
\[
\big[ B + \bar B + i \, (\bar C \, \dot C - \dot {\bar C} \, C)\big]\,\big[E \, \dot E + \frac{1}{2}\,
\{ B - \bar B - 3\, i\,(\dot {\bar C}\,C + \bar C \, \dot C)\,\} \big] = 0.
\eqno(A.7)
\]
The above equation establishes the existence of the CF-type restriction:  $B + \bar B + i \, (\bar C \, \dot C - \dot {\bar C} \, C) = 0$ 
on our theory due to the equivalence of the coupled (but equivalent) Lagrangians (i.e. $L_B \equiv L_{\bar B}$). 
This is due to the fact that, in {\it no} way, we can state that the other combination:
$E\,\dot E + \frac{1}{2}\,\{B - \bar B - 3\,i\,(\dot{\bar C}\,C + \bar C\,\dot C)\} = 0$. On the contrary, the CF-type
restriction: $B + \bar B + i \, (\bar C \, \dot C - \dot {\bar C} \, C) = 0$ has been proven from various angles [cf. Eq. (36)].

We end this Appendix with the concluding remark that we have derived the CF-type restriction on our theory from theoretical requirements related with the
symmetries of the coupled (but equivalent) Lagrangians and the absolute anticommutativity properties. However,  our present derivation of the CF-type
restriction is more {\it direct} as well as {\it transparent}.\\

\vskip 0.5cm

\begin{center}
{\bf Appendix B: On the derivation of $f_3 = E \, \dot C +  \dot E \,C$ and $f_2 = \dot{\bar B}\,C - \bar B\,\dot C$}\\
\end{center}

\vskip 0.5cm

\noindent
The theoretical content of this Appendix is, first of all, devoted to the {\it explicit} derivation of $f_3(\tau)$ in the expansion
of $\Sigma(\tau, \bar\theta)$ in the {\it anti-chiral} super expansions (41). Towards this objective in mind, we focus 
on Eq. (51) where the first-order differential equation w.r.t. the evolution parameter $\tau$ for $f_3$ has been expressed. We can
re-write it as
\[
f_3\,\dot C + {\dot f}_3\,C - \dot E\,\dot C \, C - E\,\ddot C \, C = 0 \Rightarrow 
\]  
\[
\frac{d}{d\,\tau}\,[f_3\,C] - \frac{d}{d\,\tau}\,[(\dot E\, C + E\, \dot C)\,C] = 0, \eqno (B.1)
\]
where we have used the fermionic property $(C^2 = 0)$ of the ghost variable $(C)$. The above equation can be re-expressed in a different {\it but}
useful form as the total derivative w.r.t. $\tau$:
\[
\frac{d}{d\,\tau}\,[\{f_3 - (E\,\dot C + \dot E \, C)\} \, C] = 0. \eqno (B.2)
\]
Integrating the above equation from $\tau = - \infty$ to $\tau = + \infty$ (which are the limiting cases for $\tau$ in our theory), we 
obtain the following relationship:
\[
[ f_3 - (E\,\dot C + \dot E\,C)]\, C = 0. \eqno (B.3)
\]   
We would like to point out that, while deriving (B.3) from (B.2), we have assumed that {\it all} the physical variables of the Lagrangian $L_B$ and the 
secondary variable $f_3 (\tau)$ vanish off at $\tau = \pm \, \infty $. For $C \ne 0$, we obtain the desired result: 
$f_3 = E\,\dot C + \dot E\,C$. We have taken $C \ne 0$ because the whole set of BRST symmetry transformations
in Eq. (33) is true {\it only} when the ghost variable $C(\tau)$ has the non-trivial and non-zero value.

We now concentrate on the precise determination of $f_2 (\tau)$ of the super-expansion (41). In other words, we wish to show that
$f_2 = \dot {\bar B} \, C - \bar B \, \dot C $. For this purpose, we note that we have a first-order differential equation w.r.t. the 
evolution parameter $\tau$ for the secondary variable $f_2 (\tau)$ in Eq. (49). This can be re-expressed as follows
\[
f_2\,\dot C + {\dot f}_2\,C - 2\,f_2\,\dot C - 2\, \dot {\bar B} \, \dot C \, C +  \dot {\bar B} \, \dot C \, C + \bar B \, \ddot C \, C = 0
\eqno (B.4)
\]
where we have added and subtracted $f_2 \, \dot C$ and $\dot {\bar B} \, \dot C \, C$. The above equation implies that we have now
its modified form (with total derivatives) as:
\[
\frac{d}{d\,\tau}\,[f_2\,C] - 2\,(f_2\,\dot C + \dot{\bar B}\,\dot C\,C) + \frac{d}{d\,\tau}\,[\bar B\,\dot C\,C] = 0 
\]
\[
\Rightarrow \quad \frac{d}{d\,\tau}\,\Big[f_2\,C + \bar B\,\dot C\,C\Big] - 2\,\Big[(f_2 - \dot{\bar B}\,C)\,\dot C\Big] = 0. \eqno (B.5)
\]
Using the fermionic $(C^2 = \dot C^2 = 0)$ property of the ghost variables $C$ and $\dot C$, we can recast the above equation in the 
following interesting form where $[f_2 - (\dot {\bar B} \, C - \bar B \, \dot C)]$ appears very nicely in the individual terms of the 
following difference, namely;
\[
\frac{d}{d\,\tau}\,\Big[\{f_2 - (\dot{\bar B}\,C - \bar B\,\dot C)\}\,C\Big] - 2\,\Big[\{f_2 - (\dot{\bar B}\,C - \bar B\,\dot C)\}\,\dot C\Big] = 0. \eqno(B.6)
\]
We can expand the {\it total} derivative in the {\it first} term to obtain:
\[
\frac{d}{d\,\tau}\, \Big[\{f_2 - (\dot{\bar B}\,C - \bar B\,\dot C)\} \Big]\, C - \Big[\{f_2 - (\dot{\bar B}\,C - \bar B\,\dot C)\}\,\dot C \Big] = 0. \eqno(B.7)
\] 
Defining $f_2 - (\dot{\bar B}\,C - \bar B\,\dot C) = \chi$ leads us to the following
\[
\Big(\frac{d}{d\,\tau}\,\chi\Big)\,C - \chi\,\dot C = 0 \qquad \Rightarrow \qquad \dot\chi\,C = \chi\,\dot C. \eqno(B.8)
\]
Multiplying from the right by $C$ and taking into account the fermionic (i.e. $C^2 = 0$) nature of the ghost variable $C$, we obtain the following
\[
0 = \chi\,\dot C\,C \qquad \Rightarrow \qquad \chi = 0 \;\; \; \qquad \big[\text{for} \quad C\,\dot C \ne 0\big].  \eqno(B.9)
\]
It should be noted that we have the off-shell nilpotent BRST symmetry transformation: $s_b\,C = C\,\dot C$ [cf. Eq. (33)]. As a consequence, 
the combination of the variables $C\,\dot C \ne 0$.
If the symmetry of a theory is the guiding principle behind {\it its beauty}, it is {\it physically} correct to assume that $s_b\,C = C\,\dot C \ne 0$. 
In fact, if we take $C \dot C = 0$ the whole beauty and sacrosanct properties 
(i.e. off-shell nilpotency and absolute anticommutativity) of the (anti-)BRST symmetry transformations of our present theory
will be spoiled. As a consequence, the CF-type restriction will {\it no} longer remain (anti-)BRST invariant.  It should be recalled, however,
that we have invoked the CF-type restriction in proving the equivalence of the Lagrangians [cf. Eq. (35)] w.r.t. the (anti-)BRST
symmetry transformations. Hence, 
our conclusion in (B.9) is {\it correct} which leads to the derivation of $f_2 = \dot{\bar B}\,C - \bar B\,\dot C$ from $\chi = 0$.\\

\vskip 0.3cm

\begin{center}
{\bf Appendix C: On the Proof of the Conservation Law}\\
\end{center}

\vskip 0.3cm

\noindent
We take up here the expression for the BRST charge $Q_B$ [cf. Eq. (74)] that has been derived using the Noether theorem [cf. Sec. 7]. We 
exploit the EL-EOM derived from the Lagrangian $L_B$ to recast the expression for ${\dot Q}_B$, namely;
\[
{\dot Q}_B = \frac{\dot C\,E}{2}\,(p_x^2 + 2\,m\,p_t) + \frac{C\,\dot E}{2}\,(p_x^2 + 2\,m\,p_t) + C\,E\,(p_x\,{\dot p}_x + m\,p_t) 
\] 
\[
+2\,E\,\dot E\,B\,\dot C + E^2\,B\,\dot C + E^2\,B\,\ddot C + {\dot E}^2\,B\,C + E\,\ddot E\,B\,C + E\,\dot E\,\dot B\,C
\]
\[
 + E\,\dot E\,B\,\dot C + 2\,B\,\dot B\,C - i\,\dot B\,\bar C\,\dot C\,C - i\,B\,\dot{\bar C}\,\dot C\,C 
- i\,B\,\bar C\,\ddot C\,C,  \eqno(C.1)
\]
into the following form
\[
{\dot Q}_B = -i\,E^2\,\ddot{\bar C}\,\dot C\,C - i\,E\,\dot E\,\dot{\bar C}\,\dot C\,C + 3\,E\,\dot E\,B\,\dot C + E^2\,B\,\ddot C + {\dot E}^2\,B\,C
\]
\[
+ E\,\ddot E\,B\,C + 2\,B\,\dot B\,C + B^2\,\dot C - i\,\dot B\,\bar C\,\dot C\,C - i\,B\,\dot{\bar C}\,\dot C\,C 
- i\,B\,\bar C\,\ddot C\,C, \eqno(C.2)
\]
where we have used the EL-EOM from $L_B$ as:
\[
{\dot p}_x = 0, \qquad {\dot p}_t = 0, \qquad \frac{1}{2}\,(p_x^2 + 2\,m\,p_t) = -\,\dot B\,E + i\,E\,\ddot{\bar C}\,C - i\,E\,\dot{\bar C}\,\dot C.
\eqno(C.3)
\]
The expression (C.2) can be further changed to a {\it reduced} form as follows
\[
{\dot Q}_B = i\,B\,\dot{\bar C}\,\dot C\,C - i\,\dot B\,\bar C\,\dot C\,C - i\,E\,\dot E\,\dot{\bar C}\,\dot C\,C - i\,E^2\,\ddot{\bar C}\,\dot C\,C, \eqno(C.4)
\]
if we use the EL-EOM from $L_B$ w.r.t. the variable $\bar C$ as
\[
2\,\dot B\,C + B\,\dot C + 3\,E\,\dot E\,\dot C + E^2\,\ddot C + {\dot E}^2\,C + E\,\ddot E\,C - i\,\bar C\,\ddot C\,C - 2\,i\,\dot{\bar C}\,\dot C\,C = 0.
\eqno(C.5)
\]
The expression in (C.4) can be proven to be equal to {\it zero} by using the following EL-EOM from $L_B$ w.r.t. the variable $C$, namely;
\[
i\,\dot B\,\bar C - i\,B\,\dot{\bar C} + i\,E\,\dot E\,\dot{\bar C} + i\,E^2\,\ddot{\bar C} - \ddot{\bar C}\,\bar C\,C - 2\,\dot{\bar C}\,\bar C\,\dot C
= 0. \eqno(C.6)
\]
We end this Appendix with the remark that, in exactly {\it similar} fashion, we can prove the conservation law $({\dot Q}_{\bar B} = 0)$ of the anti-BRST
charge [cf. Eq. (74)] which has been derived by exploiting the theoretical tricks of the Noether theorem.\\

\vskip 0.4cm

\noindent
{\bf Acknowledgments}\\

\noindent
Two of us (AKR and AT) gratefully acknowledge the financial support from the {\it BHU-fellowship} program of the 
Banaras Hindu University (BHU), Varanasi (U. P.),
under which, the present investigation has been carried out. Fruitful and enlightening 
comments by our esteemed Reviewer are thankfully acknowledged, too.  \\

\vskip 0.4cm

\noindent
{\bf Data Availability}\\

\noindent
No data were used to support this study.\\

\noindent
{\bf Conflicts of Interest}\\

\noindent
The authors declare that there are no conflicts of interest

\end{document}